\documentclass[prb,graphix,twocolumn,showpacs,10pts]{revtex4-1}
\usepackage[utf8x]{inputenc}
\usepackage{graphicx}
\usepackage{epsfig}
\usepackage{amsmath}
\usepackage{amsfonts}
\usepackage{amssymb}
\begin{document}
\title[]{Temperature and doping dependence of normal state spectral properties in a two-orbital model for ferropnictides}

\author{J. D. Querales Flores\footnote{Present address:
Instituto Balseiro, Universidad Nacional de Cuyo, Argentina.},$^{1,2}$  C. I. Ventura,$^{1,3}$  R. Citro$^{4}$ and  J.J. Rodr\'iguez-N\'u\~nez$^{5}$ }
\affiliation{$^{1}$Centro At\'omico Bariloche-CNEA and CONICET, Av. Bustillo 9500, R8402AGP Bariloche, Argentina}
\affiliation{$^{2}$Instituto Balseiro, Univ. Nac. de Cuyo and CNEA, 8400 Bariloche, Argentina}
\affiliation{$^{3}$Sede Andina, Univ. Nac. de R\'io Negro, 8400 Bariloche, Argentina}
\affiliation{$^{4}$Dipartimento di Fisica  ``E.R. Caianiello” and CNR-SPIN, Universit\`a degli Studi di Salerno, I-84084 Fisciano, Italy.}
\affiliation{$^{5}$Lab. SUPERCOMP, Departamento de F\'isica – FACYT, Universidad de Carabobo, 2001 Valencia, Venezuela.}

\date{\today}

\begin{abstract}

Using a second-order perturbative Green's functions approach we determined the
normal state  spectral function $A(\vec{k},\omega)$ employing a minimal model for ferropnictides.
Used before to study magnetic fluctuations and superconducting properties,  it 
includes the two effective bands related to Fe-3d orbitals proposed by S.Raghu et al. [Phys. Rev. B 77, 220503 (R) (2008)],  
and local intra- and inter-orbital correlations for the effective orbitals. 
Here, we focus on the normal state electronic properties, in particular the temperature and doping dependence of the total density of states, 
$A(\omega)$, and of $A(\vec{k},\omega)$ in different Brillouin zone regions, 
comparing them with existing angle resolved photoemission spectroscopy (ARPES) and theoretical results. 
 We obtain an asymmetric effect of electron and hole doping, quantitative agreement with the experimental chemical potential shifts,  
 as well as spectral weight redistributions near the Fermi level with temperature  
 consistent with the available experiments. In addition, we predict a non-trivial dependence of  $A(\omega)$ with temperature, 
 exhibiting clear renormalization effects by correlations. Interestingly,  investigating the origin of this predicted behaviour 
 by analyzing the evolution with temperature of the k-dependent self-energy  obtained in our approach, 
 we could identify a number of Brillouin zone points, not probed by ARPES yet,  
  where the largest non-trivial effects of temperature on the renormalization are predicted for the parent compounds.  
 
\end{abstract}


\maketitle

\section{Introduction}

In 2008, the discovery of superconductivity in  LaFeAsO$_{1-x}$F$_{x}$, with transition temperature  $T_{c} = 26 $ K at $x \gtrsim 0.05$ doping,\cite{Kamihara} 
 prompted great interest in the experimental and theoretical study  
of iron-based superconductors. LaOFeAs-like compounds, denoted the 1111-ferropnictide family, served as starting point 
and were followed rapidly by the discovery of similar electronic properties in a series of other iron-based families of compounds. 
In the meantime many new superconductors have been found, all of them including quasi-two dimensional(2D) la\-yers with a square-lattice array of Fe ions.\cite{paglione2010, zhu2009} A vast amount of experimental and theoretical effort has 
 been devoted to explore this family of compounds, evidenced by more than 15,000 articles published since 2008 according to the latest reviews, see e.g. Refs.  \cite{review2015,libro2013,review2011,Review2012} and references therein.  Therefore, in the following we will focus on the previous research work 
 most relevant to provide a proper context  for the discussion of the open problem which we have addressed, and the new results 
 on the normal state spectral properties we obtained.

It is generally recognised that the ground state properties of the ferropnictide parent compounds are 
mostly  well described by first principles density functional theory (DFT) calculations, 
 indicating  that the low energy excitations are mainly due to electrons in the Fe-3d orbitals.\cite{Lebeque2007,Kamihara2006}
 For LaOFeAs, calculations were carried out almost simultaneously in a number of works.\cite{singh2008,craco2008,mazin2008,singh2008-2,singh2009}
The Fermi surface topology and band structure are rather similar for the 11 (e.g. FeSe), 111 ( e.g. LiFeAs, NaFeAs ), 122 ( e.g. BaFe$_{2}$As$_{2}$, CaFe$_{2}$As$_{2}$, 
EuFe$_{2}$As$_{2}$), 1111 ( e.g. LaOFeAs, LaFePO, SmFeAsO,CeFeAsO) series of iron-based superconductors.\cite{yang2010, YangPRL2009, He2010, Zhang2010, Chen2010} 
However, as a function of doping the agreement between DFT calculations
and ARPES experiments  in materials like LaOFeAs  diminishes,\cite{mazin2008} and unexpected Fermi surface topology changes were found as
a function  of temperature and doping.\cite{ding2011, holder2012, chang2011, cui2012}

Theoretical predictions showed that the 1111 compounds consist of two quasi-2D Fermi cylinders at the zone center ($\Gamma$) and a massive 3D hole pocket at the Z-point.\cite{singh2008, nekrasov2008} This behavior was observed by ARPES  and quantum oscillation measurements.\cite{coldea2008} On the other hand,
in the case of 122 compounds, observations related to the dimensionality of the electronic structure revealed quite different behavior, i.e. a more 3D nature of the electronic
structure was found in electron doped BaFe$_{2}$As$_{2}$ \cite{vilmercati2009,thirupathaiah2010}  whereas a quasi-2D electronic structure was derived for the case of K-doped BaFe$_{2}$As$_{2}$.\cite{Zabolotnyy2009}
 Interestingly, Lui et al.\cite{liu2009} showed a transformation of the nature of the electronic structure of CaFe$_{2}$As$_{2}$ from quasi-2D to more 3D 
as a function of temperature, going from the high-temperature tetragonal to the low-temperature orthorhombic phases. 

Many aspects of the temperature dependence of the normal state spectral properties remain largely unstudied in ferropnictides, 
which motivated our present work.  Here, we present our study of the changes with temperature and doping of the normal state 
electronic structure,  throughout  the Brillouin zone (BZ). We  compare our results with ARPES experimental and previous theoretical results, where feasible, 
and  have been able  to predict interesting  non-trivial temperature dependent effects at Brillouin zone points, yet unexplored by ARPES.
To do this, we  employed  a minimal microscopic model, which includes two  effective bands\cite{raghu} to describe the low energy band structure, 
 as well as intra- and inter-orbital local Coulomb interactions, as detailed in next section. 
 Similar effective two-orbital models have been studied before, focusing on other aspects of the problem,  mostly on the superconducting properties.  
 For example, investigations of pairing mechanisms and gap symmetry were reported,\cite{Graser2009,yao2009,Ran2009,zhou2011,Review2011,jhu2012,gliu2014,ptok2015} studies of 
spin fluctuations\cite{Graser2009,yao2009,gliu2014} and a spin-density wave phase,\cite{eremin2009} 
Mott transition for strong electron correlations,\cite{qsi1,qsi2,yamada2014} lattice and orbital properties,\cite{calderon2009,aperis2011,liu2011,yamada2014} etc.
 In our case,  we studied the minimal microscopic two-orbital model  using analytical perturbative techniques to take into account the effect of electron correlations,
in order to determine the relevant normal state Green's functions and the corresponding  temperature-dependent electronic spectral density. 
We would like to highlight the fact that, in contrast to other analytical techniques,   
 in our approach  a k-dependent  self-energy  is obtained, also dependent on temperature and doping, 
 enabling us to explore normal state physical properties throughout the Brillouin zone, and thus address new problems. 
 
Our perturbative  treatment  for the electronic correlations in ferropnictides is justified by previous estimations of intermediate values for them, 
both from  numerical calculations and experimental results. 
For example, a combination of local density approximation for DFT and dynamical mean field theory (DMFT) calculations 
 for REOFeAs (RE = La, Ce, Pr, and Nd),\cite{miyake2008} estimated a Hubbard local intra-orbital correlation value: $U \sim 3.69$ eV. 
 Refs.\cite{haule2008,craco2008} estimated that with  $U$ = 4 eV, width of Fe-bands $W_{Fe} = 3$ eV from LDA, and a Hund′s
coupling energy $J$ = 0.7 eV,  taking for the  inter-orbital correlation $ V = U - 2 J $ as in most refs.,\cite{miyake2008,craco2009,shorikov2009,craco2011,Skornyakov, Skornyakov2011, Skornyakov2012} 
in this case yielding $ V = 2.6 eV$,  the available normal state ARPES results could be described, 
while the same parameters were also adopted for studying  the superconducting state in Ref.\cite{craco2009}. 
Furthermore, it was concluded\cite{haule2008}  that  $U$ = 4.5 eV would transform the
system into a Mott-insulator, in contrast to other LDA+DMFT predictions,\cite{Skornyakov, Skornyakov2011, Skornyakov2012, shorikov2009} reporting that the system remains metallic  and does not transform into a Mott-insulator even increasing $U$ up to 5 eV, with $W_{Fe} = 4 $eV and $J = 0.7$ eV;
while for $ J = 0$ an electronic structure characteristic of much weaker correlations is obtained, with a quasiparticle weight renormalization factor 
$Z \sim 0.8$.\cite{shorikov2009} From ARPES, mass renormalization factors between 1.5 and 2.5  were estimated in 122 compounds of the form Sr$_{1-x}$K$_{x}$Fe$_{2}$As$_{2}$, depending on the Fermi surface  sheet,\cite{Yoshida, Lu2008, Liu2008, Terashima} with similar reports of 1.3 - 2.1 for 1111 and  other 122 compounds.\cite{Suchitra2008, coldea2008, Analytis2009}  X-ray absorption
spectroscopy  estimations  of U, placed it below $\sim$ 4 eV.\cite{yang2009, kroll2008} To describe the smaller 
 experimental Fermi-surface areas reported in de Haas- van Alphen experiments in SrFe$_{2}$As$_{2}$,\cite{Suchitra2008} 
 a renormalization of the LDA band structure was suggested in 2009,\cite{cappelluti-2009} estimating an interband scattering of magnitude $\sim 0.46 eV$ between hole and electron bands  to explain the reported experimental band shifts.\cite{coldea2008}    
 Regarding the minimal two-orbital model  by Raghu et al. of Ref.\cite{raghu}, the full bandwidth of the non-interacting band structure is about 12 eV, 
 and  local intra-orbital correlation values $ U / W = 0.2 - 0.5 $ have been used previously.\cite{yao2009,Graser2009} 
 Adding Hund's coupling $J$ and local Coulomb correlations, a Mott  transition in the two-orbital model\cite{raghu} was predicted at a critical interaction $  U_{c} / W \sim 2.66/(1+J/U)$,\cite{qsi2} if $J/U < 0.01$, with a decrease of $U_{c}$ for larger $J$: deducing that the effect of increasing Hund's coupling 
 on critical $U_{c}$ is similar to the effect of having more degenerate orbitals in a multiorbital Hubbard model,\cite{qsi2} effectively weakening the effect of correlations \cite{shorikov2009} and stabilizing a metallic state.\cite{qsi2}    

In addition to effective two-orbital models for ferropnictides, other multiorbital effective models were proposed.\cite{review2015} 
A  three-orbital Hamiltonian was constructed involving the $3d$ orbital $xz$,$yz$ and $xy$ for Fe\cite{daghofer2010}, and compared 
with the two-orbital model of Ref.\cite{raghu}.  Improvement of two shortcomings of the latter were reported,  related to the relative weights 
of each orbital on the Fermi surface. Also effective four-orbital \cite{qsi2}  and five-orbital models\cite{Graser2009,craco2008,craco2009,craco2011,review2015,arita2008} were studied. Nevertheless, the effective two-orbital model of Ref.\cite{raghu} is still recognized as a useful minimal model to describe the main features of the low-energy physics of ferropnictides.\cite{gliu2014,yamada2014,jhu2012,qsi2,Graser2009,yao2009,Ran2009,Hu2009,zhou2011,Review2011}

     Our paper is organized as follows. In  Section \ref{model} we present the microscopic correlated two-orbital model adopted for our study of the normal state 
     of Fe-based superconductors, and describe the analytical Green's function approach we used to calculate the electronic spectral density function 
     and the total density of states including the correlations in second-order of perturbations ( further details of our analytical calculations 
     appear in Supplementary Appendix A).
      In Section \ref{results} we present and discuss electronic structure results obtained at different temperature and doping values along the Brillouin zone, 
      and compare them with available theoretical and experimental ARPES results for the normal state spectral properties of ferropnictides. 
      To understand the origin of the non-trivial temperature dependence we predict for the density of states,  in Supplementary Appendix B  we  
       analyze the k-dependent electron self-energies obtained in our approach, having  identified a number of specific Brillouin zone points, 
       not probed by ARPES experiments yet, where we find the largest non-trivial effects of temperature on the renormalization.  
    We conclude in Section \ref{conclusions}, summarizing our main findings with the minimal model for pnictides 
    and analytical treatment used, prompting for new ARPES experiments which could test our predictions, 
    and mentioning possible extensions and applications of our work.

\section{Microscopic model and analytical approach.}\label{section2}

\subsection{Correlated effective two-orbital model.}
\label{model}

To describe analitically the properties of ferropnictides, we will consider a simplified model which contains the minimum number of
degrees of freedom  preserving the essential physics of the problem in 1111 and 122 compounds, as mentioned in previous section. 
In particular, a minimal two-orbital model suitable to describe the Fermi surface topology 
of ferropnictides was proposed by Raghu et al. in Ref.\cite{raghu}. The model  consists of a two-dimensional square lattice,
 with each site having two degenerate orbitals. Raghu et al. in Ref.\cite{raghu} fitted the bare-band tight-binding parameters, 
 to obtain an effective band structure which, after folding the Fermi surface  to the two Fe/cell Brillouin zone, 
 exhibits two hole pockets around the  $\Gamma$  point and two electron pockets around the $M$ point, which are characteristic of the majority of FeAs compounds.\cite{YangPRL2009,yang2010,He2010,Zhang2010,Chen2010} 
 On the large one Fe/cell Brillouin zone, which we will be using in this work, 
 the unfolded Fermi surface includes a hole-related band, relevant to the description of the physics near the $\Gamma$ point, 
 and an electron-related band to describe the physics near the $M$ point. The two-orbital model was also  shown to be suitable
   to describe the  extended s-wave pairing  believed to be relevant for ferropnictides, and other details of the superconducting phase.\cite{Graser2009,yao2009,Ran2009,Hu2009,zhou2011,Review2011,jhu2012,gliu2014}

  In this work, we will study the doping and temperature effects on the electronic structure and spectral properties, using the following 
  microscopic model for the normal state of pnictides: 
\begin{equation}
\mathcal{H}  = \mathcal{H}_{0} + V_{int}.
\end{equation}

Thus, we will be modelling these compounds by an effective extended Hubbard model: consisting of two correlated electronic orbitals, described by 
Hamiltonian $\mathcal{H}_{0}$ including the two effective bands  proposed by Raghu et al.\cite{raghu}  in the large BZ with one Fe atom/unit cell, 
and local intra- and interorbital electronic correlations 
included in  $V_{int}$. Concretely: 

\begin{eqnarray}
 {\mathcal{H}}_{0} &  = &\sum_{k,\sigma}{\left[{E}_{c}(k) {c}^{\dagger}_{k\sigma}{c}_{k\sigma} + {E}_{d}(k) {d}^{\dagger}_{k\sigma}{d}_{k\sigma}\right]}
  \label{Hamiltonian0}
\end{eqnarray}

\noindent where the operators ${c}^{\dagger}_{k \sigma}$ and $d^{\dagger}_{k \sigma}$ create  an electron in the respective 
$c$ and $d$  effective bands, with spin $\sigma= \uparrow, \downarrow$ and crystal momentum $\vec{k}$ while,  following Ref.\cite{raghu},  
we consider for the two bare effective bands:

 \begin{equation}
  E_{c,d}(\vec{k}) = \epsilon_{+}(\vec{k}) \pm \sqrt{ \epsilon_{-}^{2}(\vec{k}) + \epsilon_{xy}^{2}(\vec{k})} -\mu
  \label{effectivebands}
 \end{equation}

\noindent $\mu$ denotes the chemical potential: notice that we refer the energies to $\mu$, since we will adopt the grand canonical ensemble 
in our Green's function treatment of the system, like  in  Ref.~\cite{raghu}.  
$\sigma$ denotes the spin degrees of freedom, and

 \begin{eqnarray}
  \epsilon_{\pm}(\vec{k})&=& \frac{\epsilon_{x}(\vec{k}) \pm \epsilon_{y}(\vec{k}) }{2}   \\
  \epsilon_{xy}(\vec{k}) & = & -4t_{4}\sin(k_{x})\sin(k_{y}) \nonumber \\
  \epsilon_{x}(\vec{k}) & = & -2t_{1}\cos(k_{x}) - 2t_{2}\cos(k_{y})- 4t_{3}\cos(k_{x})\cos(k_{y})  \nonumber \\
  \epsilon_{y}(\vec{k}) & = & -2t_{2}\cos(k_{x}) - 2t_{1}\cos(k_{y}) -4t_{3}\cos(k_{x})\cos(k_{y}) \nonumber
\end{eqnarray}

\noindent where the parameters $t_{i}, i=1,4, $ denote the hopping amplitudes between sites on the  square lattice formed by the Fe atoms\cite{raghu} in ferropnictides.
                                                                                                                                                     
As for the electronic correlations,  we assume that short-range Hubbard-like electron-electron interactions are present,  and consider two local correlations between the effective orbitals at each site ( as in Refs. \cite{raghu,yao2009}):    intra-orbital Coulomb repulsion $U$, and  inter-orbital repulsion $V$.
 Therefore $V_{int}$ has  the following form: 

 \begin{align}
& {V}_{int}  =  \sum_{i}\left[ U \left( {n}_{i\uparrow}{n}_{i\downarrow} +{N}_{i\uparrow}{N}_{i\downarrow} \right)  + V \left({n}_{i\uparrow}+{n}_{i\downarrow}\right) \left({N}_{i\uparrow}+{N}_{i\downarrow}\right) \right]
\raisetag{1pt}
 \label{Vint}
 \end{align}

Here: $n_{i\sigma} = {c}^{\dagger}_{i\sigma}{c}_{i\sigma}$ and $N_{i\sigma} = {d}^{\dagger}_{i\sigma}{d}_{i\sigma}$, and $i$ denotes the lattice sites.
Fourier transforming, $V_{int}$ can be written in $k$-space as follows:

\begin{align}
 V_{int}  =  \frac{U}{N} \sum_{k_{1},k_{2},k_{3}} \left[  {c}^{\dagger}_{k_{1}\uparrow}{c}_{k_{2}\uparrow}  {c}^{\dagger}_{k_{3}\downarrow}{c}_{k_{1}-k_{2}+k_{3}\downarrow}  + {d}^{\dagger}_{k_{1}\uparrow}{d}_{k_{2}\uparrow}  {d}^{\dagger}_{k_{3}\downarrow}{d}_{k_{1}-k_{2}+k_{3}\downarrow}  \right] \nonumber \\ 
           +    \frac{V}{N} \sum_{k_{1},k_{2},k_{3}} \left[  {c}^{\dagger}_{k_{1}\uparrow}{c}_{k_{2}\uparrow}  {d}^{\dagger}_{k_{3}\uparrow}{d}_{k_{1}-k_{2}+k_{3}\uparrow} + {c}^{\dagger}_{k_{1}\uparrow}{c}_{k_{2}\uparrow}  {d}^{\dagger}_{k_{3}\downarrow}{d}_{k_{1}-k_{2}+k_{3}\downarrow} \right]  \nonumber \\ 
	   +   \frac{V}{N} \sum_{k_{1},k_{2},k_{3}} \left[  {c}^{\dagger}_{k_{1}\downarrow}{c}_{k_{2}\downarrow}  {d}^{\dagger}_{k_{3}\uparrow}{d}_{k_{1}-k_{2}+k_{3}\uparrow} + {c}^{\dagger}_{k_{1}\downarrow}{c}_{k_{2}\downarrow}  {d}^{\dagger}_{k_{3}\downarrow}{d}_{k_{1}-k_{2}+k_{3}\downarrow} \right]
\raisetag{1.5pt}
\end{align}

\subsection{Our electronic structure calculation for ferropnictides.}
\label{calculations}

The total electron  spectral density  to be compared with ARPES experiments is  $A(\vec{k},\omega)$, which 
in our effective extended Hubbard model is defined by:   
\begin{equation}\label{partial}
A(\vec{k},\omega) = A_{c}(\vec{k},\omega) + A_{d}(\vec{k},\omega)
\end{equation}
with contributions from both electron bands given by: 

\begin{equation}
 A_{c\sigma}(\vec{k},\omega)  =  -\frac{1}{\pi} Im G^{ret}_{\sigma}(k,\omega) ;  \, \, \, \, \, A_{d\sigma}(\vec{k},\omega)  =  -\frac{1}{\pi} Im F^{ret}_{\sigma}(k,\omega)
 \label{partialdos}
 \end{equation}

  Here: $G^{ret}_{\sigma}(k,\omega) $ and $F^{ret}_{\sigma}(k,\omega) $ denote the retarded Green's functions corresponding  
the  $c$ and $d$ electrons in our two effective bands, i.e.  
\begin{align}\label{g-retarded}
G_{\sigma}^{ret}(\vec{k},\omega)  = G(k,\omega + i\delta)  = \ll c_{k \sigma};c^{\dagger}_{k \sigma}\gg (\omega + i\delta) 
\end{align} 
\begin{align}\label{f-retarded}
F_{\sigma}^{ret}(\vec{k},\omega)  = F(k,\omega + i\delta)  = \ll d_{k \sigma};d^{\dagger}_{k \sigma}\gg(\omega + i\delta)
\raisetag{1pt}
\end{align} 
where $\delta$ is an infinitesimal positive number. Integrating the spectral density over the 1st Brillouin zone,  yields the total density of states (TDOS), or local spectral function: 
\begin{equation} 
A(\omega) = \sum_{k \in BZ} \,  A(k,\omega)  \quad 
\label{totalA}  
\end{equation}

We calculated the temperature-dependent Green's functions using Zubarev's equations of motion (EOM) formalism,\cite{zubarev,nolting} i.e.:
\begin{equation}
 \omega  \ll \widehat{A};\widehat{B}  \gg = \frac{1}{2\pi} \langle \{\widehat{A},\widehat{B}\}\rangle + \ll [\widehat{A},\widehat{\mathcal{H}}]; \widehat{B}  \gg (\omega)
 \nonumber
\end{equation}
 \noindent  where $\widehat{A}$ and $\widehat{B}$ are fermionic operators,
 $\ll \widehat{A};\widehat{B}  \gg$ is the time Fourier-transform  of the retarded Green's function
$-i\theta(t-t')\langle  \widehat{A}(t)\widehat{B}(t') + \widehat{B}(t')\widehat{A}(t) \rangle$,  where the time-dependent operators appear in Heisenberg representation,  
and the expectation value is calculated using the appropriate statistical ensemble at finite temperature $T$ (or  the ground state of the system, at T=0 )\cite{zubarev}. 
In our case, we study the normal state of the system at temperature $T$, and in particular evaluate
 the required expectation values in the paramagnetic normal phase, using the grand canonical ensemble.           

We obtained 
the following exact  coupled set of equations of motion  for  $G_{\sigma}(k,\omega)$ and $F_{\sigma}(k,\omega)$, respectively:

\begin{align}
 & \left[\omega - E_{c}(k) \right] G_{\sigma}(k,\omega)  =  \frac{1}{2\pi} + \sum_{k_{1},k_{2}}\left[\frac{U}{N} {\Gamma}_{1}(k_{1},k_{2},k,\omega)   \right.  \nonumber \\ 
 &\left.   + \frac{V}{N} {\Gamma}_{2} (k_{1},k_{2},k,\omega) + \frac{V}{N} {\Gamma}_{3} (k_{1},k_{2},k,\omega) \right]    \label{eqG}\\
 &\left[\omega - E_{d}(k) \right] F_{\sigma}(k,\omega)  =  \frac{1}{2\pi} +\sum_{k_{1},k_{2}}\left[\frac{U}{N} {\Gamma}_{4}(k_{1},k_{2},k,\omega)  \right. \nonumber \\ 
  & \left. + \frac{V}{N} {\Gamma}_{5} (k_{1},k_{2},k,\omega) + \frac{V}{N} {\Gamma}_{6} (k_{1},k_{2},k,\omega) \right]   \label{eqF} 
\raisetag{1pt}
\end{align}

 \noindent where in Eq. \ref{eqG} we denoted:   

\begin{eqnarray}\label{set1}
 {\Gamma}_{1}(k_{1},k_{2},k,\omega) & \equiv & {\Gamma}^{ccc}_{\sigma, \overline{\sigma},\overline{\sigma}}(k_{1},k_{2},k,\omega) =  \nonumber \\ 
& = & \ll c_{k_{2}, \sigma}  c^{\dagger}_{k_{1}, \overline{\sigma}}  c_{k_{1}-k_{2}+k, \overline{\sigma}}  ;  c^{\dagger}_{k \sigma} \gg (\omega)\nonumber \\ 
{\Gamma}_{2}(k_{1},k_{2},k,\omega) & \equiv & {\Gamma}^{cdd}_{\sigma, \sigma ,\sigma}(k_{1},k_{2},k,\omega) \nonumber \\
{\Gamma}_{3}(k_{1},k_{2},k,\omega) & \equiv & {\Gamma}^{cdd}_{\sigma, \overline{\sigma} ,\overline{\sigma}}(k_{1},k_{2},k,\omega)
\end{eqnarray}

 \noindent and in Eq. \ref{eqF}: 
\begin{eqnarray}\label{set2}
 {\Gamma}_{4}(k_{1},k_{2},k,\omega) & \equiv & {\Gamma}^{ddd}_{\sigma, \overline{\sigma},\overline{\sigma}}(k_{1},k_{2},k,\omega) \nonumber \\
{\Gamma}_{5}(k_{1},k_{2},k,\omega) & \equiv & {\Gamma}^{dcc}_{\sigma, \sigma ,\sigma}(k_{1},k_{2},k,\omega) \nonumber \\
{\Gamma}_{6}(k_{1},k_{2},k,\omega) & \equiv & {\Gamma}^{dcc}_{\sigma, \overline{\sigma} ,\overline{\sigma}}(k_{1},k_{2},k,\omega)
\end{eqnarray}

\noindent having introduced the following notation, above, for simplicity:

\begin{equation}\label{definition}
  \Gamma^{\alpha,\beta,\gamma}_{\sigma_{\alpha}\sigma_{\beta}\sigma_{\gamma}}(k_{1},k_{2},k,\omega) \equiv 
    \, \,  \ll c_{\alpha_{k_{2}, \sigma_{\alpha}}} c^{\dagger}_{\beta_{ k_{1}, \sigma_{\beta}}} c_{\gamma_{k_{1}-k_{2}+k,\sigma_{\gamma}}}  ;  c^{\dagger}_{k \sigma}\gg (\omega)\\ 
\end{equation}

\noindent where $c_\alpha, c_\beta, c_\gamma$ describe c- or d-annihilation operators and $ \sigma_{\alpha}, \sigma_{\beta},\sigma_{\gamma} $ their spins ($ \sigma$, or $ {\overline{\sigma}}$), as required by each equation  (more details in supplementary Appendix A).

 Notice that the Hartree-Fock solution to this problem would be obtained by closing this set of equations of motion in first-order, 
with a mean-field approximation   of $\Gamma_{i}$ $(i = 1,6)$ in terms of  $G_{\sigma}(k,\omega)$ and $F_{\sigma}(k,\omega)$, as detailed in supplementary Appendix A.

In our  work, instead, we calculated the equations of motion for the three $\Gamma_{i}$ $(i = 1,3)$ functions  coupled to  $G(\vec{k},\omega)$ in first order 
as stated in the previous equations; and proceeded likewise for $\Gamma_{i}$ $(i = 4,6)$ coupled to $F(\vec{k},\omega)$. 
Each of these second-order equations of motion introduces three new coupled  higher-order Green's functions in the problem.

To close and solve the coupled set of equations of motion in second-order, we used the following approximation:
all higher-order Green's functions introduced in each subset of equations of motion  for $\Gamma_{i}$ $(i = 1,3)$, were approximated  in mean-field 
in terms of $\Gamma_{i}$ $(i = 1,3)$  and $G(\vec{k},\omega)$ (introducing proper average values), yielding a $4\times4$ closed set of equations of motion ( for details, see supplementary Appendix A). 
We proceeded likewise for the subset of equations for  $\Gamma_{i}$ $(i = 4,6)$ related to  $F(\vec{k},\omega)$. 
Notice that this RPA-like approximation leads to a second-order closed system of eqs. of  motion for $G$,  $\Gamma_{1}$, $\Gamma_{2}$, $\Gamma_{3}$; 
and an analogous one for $F$,  $\Gamma_{4}$, $\Gamma_{5}$, $\Gamma_{6}$; which after lengthy calculations we could solve. In supplementary Appendix A we detail the  full 
expressions obtained for these Green's functions, which might be used to calculate other normal state properties,  and in supplementary Appendix B
we discuss the corresponding k-dependent electron self-energies, which are obtained in our level of approximation.

The decoupling scheme proposed for the higher order Green's functions, which appear coupled in the EOM, accounts for the self-energy effects of the single-particle Green's function. These self-energy effects may describe both intra- and inter-orbital coupling with equal and opposite spins at lowest order (charge and spin fluctuations), thus leading to band splitting effects, which are not a mere renormalization of the energy spectrum. This fact is well known for other Hubbard-like models (see e.g. Refs. \cite{roth1969,nolting}), and the band splitting effects 
may lead to a metal-insulator transition or to charge and spin instabilities. 
 
 We shall be interested in describing the behavior of this many-particle system at finite temperatures. For a system in thermodynamic
equilibrium the expectation value of any operator may be computed by using the grand-canonical ensemble.  
To complete our solution in RPA approximation, the total electron band filling $n$ (or the chemical potential $\mu$) at temperature $T$ is determined self-consistently by the following equation:

\begin{align}\label{fill}
  n (\mu) = \int \frac{d\vec{k}}{(2\pi)^{2}} \frac{d\omega}{2\pi} \left[ A_{c}(\vec{k},\omega) \frac{1}{e^{\beta  E_{c}(\vec{k})} +1 }  \right. 
    \left. + A_{d}(\vec{k},\omega) \frac{1}{e^{\beta E_{d}(\vec{k})} +1 } \right]
\raisetag{1.5pt}
\end{align}

Thus, a major self-consistent set of coupled equations has to be solved, since the total spectral density function, which we obtain 
by solving two coupled sets of Green's functions, is required and will be integrated over the Brillouin zone ($\vec{k}$) and all  
energies $\omega$, weighed by the Fermi occupation factors  of each effective band at temperature $T$ .
The thermodynamic state of the system is defined by the parameter $\mu$ and $\beta$, the inverse temperature measured in energy units, i.e.     $\beta = \frac{1}{k_{B}T}$, where $k_{B}$ is Boltzmann's constant. Zero temperature, or $\beta \rightarrow \infty$, describes the ground state of the system. Notice that in this model, half-filling ($ n = 2 $) 
corresponds to the ferro\-pnic\-tide parent compounds.

As detailed in supplementary Appendix A, to evaluate the Green's functions $G_{\sigma}(k,\omega)$ and $F_{\sigma}(k,\omega)$,  
we need to perform double and triple summations over the first Brillouin zone of the crystal lattice.
For simplicity, we have assumed a square lattice, as done previously,\cite{raghu}  and have performed the BZ summations 
using the Chadi-Cohen BZ sampling method.\cite{macot}

\section{Results and discussion} 
\label{results}

Below, we present electronic structure results at different temperature and doping values, which we obtained for the normal
state of ferropnictides with the extended Hubbard model and our analytical treatment presented in previous section, 
and compare them with available theoretical and  experimental ARPES results in ferropnictides.
 The self-consistent sets of equations obtained for the Green's functions  
detailed in previous section, were solved numerically using the following tight-binding interaction parameters: $ t_{1} = -1 $ eV,
$t_{2}  = 1.3 $ eV , $t_{3}$  = $t_{4}  = -0.85 $ eV,  adopted in Ref. \cite{raghu}  for the two effective bands in 
the large one Fe/cell Brillouin zone:  BZ which was also  used  throughout  our work.  
In the figure captions we specify the 
precision used in the Chadi-Cohen sampling method \cite{macot} for Brillouin zone summations, including the order $\nu$ used. 
We found good convergence in most results using the seventh order ($\nu=7$) of the 
Chadi-Cohen method,\cite{macot} but the spectral function and total density of states results here presented 
were obtained using ninth or eighth order, respectively, for improved accuracy. Note that $\nu= 8$-th order for the square-lattice BZ sampling 
implies using 8,256 special BZ points, while $\nu = 9$ implies using 32,896 ones.\cite{macot}

We will start by analizing in subsection \ref{UV-effect} the effect of correlations in our model, 
comparing our calculations of the density of states  
to previous known results, like experimental data now available for 1111 and 122 compounds, 
in order to obtain a qualitative estimation of the relevant magnitudes  of $U$ and $V$ in the model. 

Next, in subsection \ref{k-effect} we exhibit  the momentum 
dependence of the spectral density function along symmetry paths of the square lattice Brillouin zone (BZ), 
comparing our results with reported ARPES data. Notice that in order to allow for direct comparison 
with the published ARPES intensity results, the same treatment is adopted  here for the data: 
i.e.  all spectral density data here presented ( denoted $ \tilde{A} ( \vec{k}, \omega)$, here) have been convoluted 
with an energy resolution of 30 meV, and also multiplied by the Fermi-Dirac function $f_{FD}(\omega)$ at temperature $T$ 
as in the experimental references cited.

Subsequently, in subsection  \ref{doping-effect} we study the effects of hole and of electron doping in terms of our model, and compare them 
with the electronic properties of ferropnictides observed in ARPES. Finally, in subsection  \ref{T-effect} we study and compare with available experimental data  
the temperature dependence of the total density of states and the spectral density at the most relevant BZ points.

\subsection{Effect of intra- and inter-orbital Coulomb interactions on the total density of states $A(\omega)$}      
 \label{UV-effect}

As discussed in the Introduction, previous theoretical and  experimental research works estimate                 
intermediate values for the electronic correlations, i.e. most of them placing  the local intra-orbital Hubbard repulsion 
in the range  $U \sim 3.69$ eV for ReFeAsO (Re= Ce, Pr, Nd)\cite{miyake2008} to  $U$ = 4.5 eV for LaO$_{1-x}$F$_{x}$FeAs,\cite{haule2008,shorikov2009}  while X-ray absorption
experiments suggest $U \leq$ 4 eV for LaO$_{1-x}$F$_{x}$FeAs,\cite{yang2009}
and BaFe$_{2}$As$_{2}$.\cite{kroll2008} For the inter-orbital correlation $V$ similar values are estimated in calculations  
for Fe-Se compounds.\cite{craco2011}                                         

\begin{figure}[t!]
  \begin{center}
  \includegraphics[width=7.5cm]{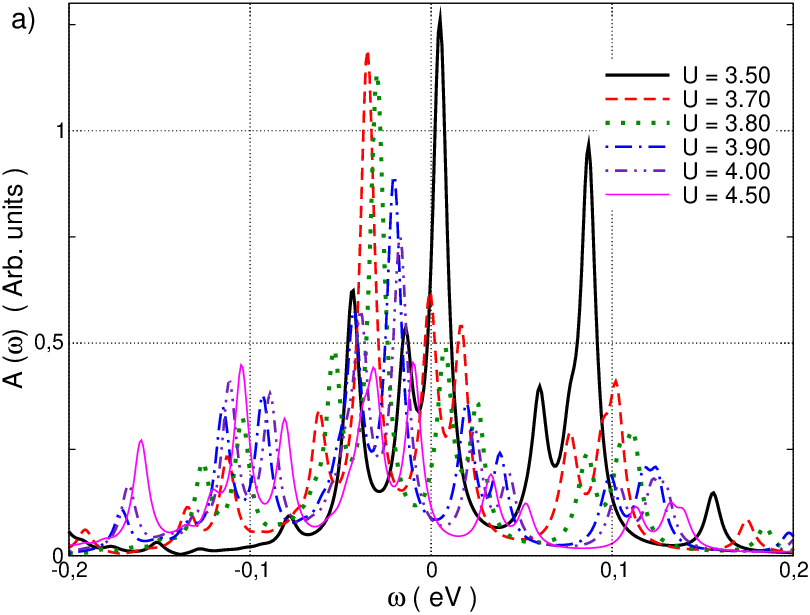}
   \includegraphics[width=7.5cm]{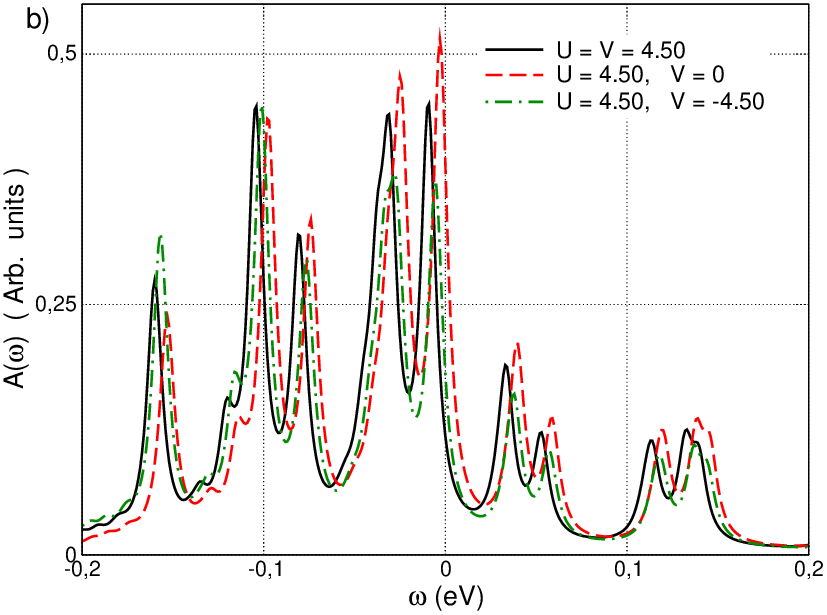}
    \caption[]{a) Dependence of total DOS $A (\omega)$ on the intra-orbital repulsion U, for fixed $V= 3.5eV $; b) Effect of the inter-orbital local Coulomb interaction V on the total DOS, for fixed $U= 4.5eV $. At half-filling: $n  = 2$, and temperature: T = 20 K; and as in Ref. \cite{raghu}: $t_{1}$ = −1.0, $t_{2}$  = 1.3, $t_{3}$  = $t_{4}$  = − 0.85 (in eV). Chadi-Cohen BZ summations order: $\nu= 8$. Note that $\omega$ is measured w.r. to $\mu$, in our treatment.  All energies are given in eV, henceforth.}
  \label{dep-U}
  \end{center}
  \end{figure}

We analized the separate effects of the intra- and inter-orbital correlation parameters included in the model. 
First,   the total density of states  $ A(\omega) $, is shown in Fig.  \ref{dep-U}(a). 
for different values of intra-orbital $U$ and a fixed inter-orbital: $V = 3.5 eV$, at half-filling $n = 2$ and temperature $T = 20$K. 
We find a large spectral weight reduction at the Fermi level in the range $U = 3.7 - 3.9$ eV, 
 with an important spectral weight redistribution around the Fermi level. 
 
In Fig. \ref{dep-U}(b), we focus on the effect of the inter-orbital Coulomb
 interaction $V$ on the density of states. Note that the TDOS is almost independent of $V$, in the range of $U$ values analized. 
 We thus confirm that the most relevant Coulomb correlation in 1111 and 122 systems is in fact the local intra-orbital repulsion $U$, 
 in agreement with previous research work.\cite{Review2011, Review2012}  
   
In the following, we will focus on the dependence of the electronic structure on crystal momentum, doping and  temperature, 
fixing the local Coulomb interactions as: $U = V \sim 3.50$ eV, typical  intermediate correlation values  characteristic of Fe-pnictide compounds 
as previously discussed.\cite{yang2009,shorikov2009,kroll2008,kurmaev2008} Being the total bandwidth $W \sim 12 $ eV for the non-interacting effective band structure 
of Ref.\cite{raghu}, the value $ U = 3.5 $ eV corresponds approximately to $ 0.29 W $.

\subsection{Momentum dependence of the spectral density $\tilde{A}(\vec{k},\omega)$}
 \label{k-effect}     

\begin{figure}[h!]
  \begin{center}
  \includegraphics[width=7cm]{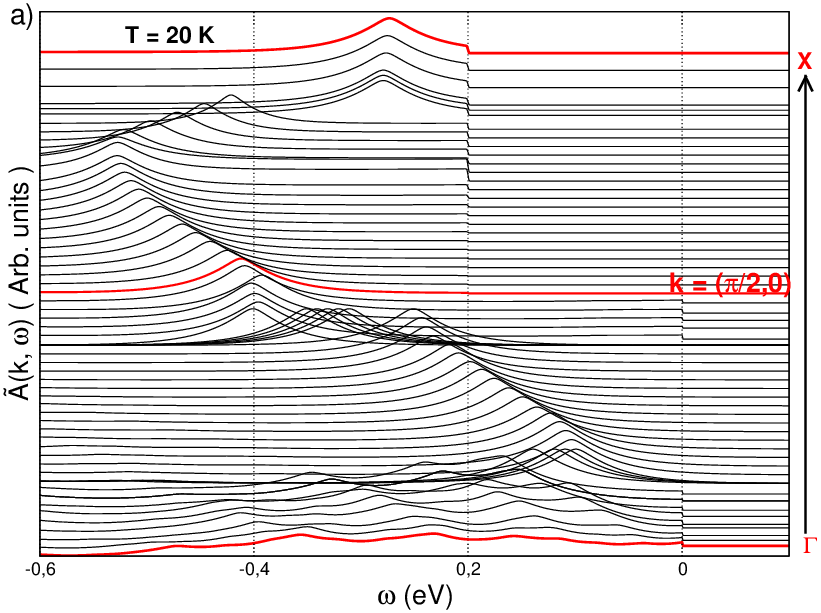}
\includegraphics[width=7cm]{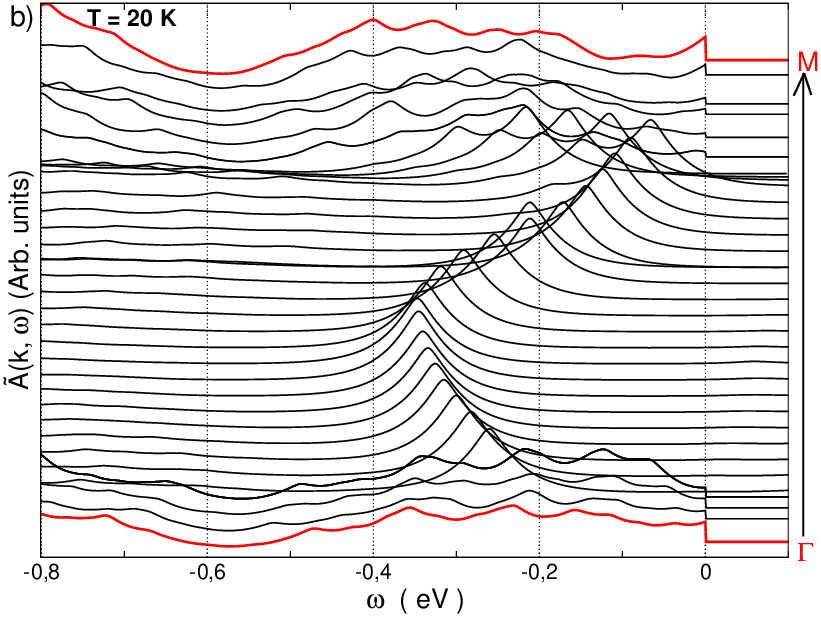}
    \caption[]{ a) Spectral density $ \tilde{A} (\vec{k},\omega)$ (i.e. convoluted with an exp. resolution of 30 meV, and multiplied by $f_{FD}(\omega)$,  as in expts.) as a function of energy $\omega$, for different BZ points 
    (shown vertically displaced)
     along the symmetry path from $\Gamma$ (bottom curve) to X (top curve).  
     b) $   \tilde{A} (\vec{k},\omega)$ as a function of $\omega$, for different BZ points along $\Gamma$-M. At T = 20 K, and $ n = 2 $ (as in Ref.\cite{Lu2009}). Here: $U = V = 3.5   eV $; $\mu = 0.15 eV$.  
    Chadi-Cohen BZ summations order: $\nu= 9$. Other parameters as in Fig.~\ref{dep-U}.
                   }
  \label{camino-GX}
  \end{center}
 \end{figure}

Our analysis of the two-orbital extended Hubbard
model included the calculation of the one-particle spectral function. In this section,
we will show our electronic structure results along two BZ paths, namely along $\Gamma$ - $X$ in Fig. \ref{camino-GX}(a),
 and along $\Gamma$ - $M$  
in Fig. \ref{camino-GX}(b), where: $\Gamma = (0, 0)$, $X = (\pi, 0)$ and $M = (\pi, \pi)$. 
These paths were chosen as they allow us to compare our results with available ARPES data, in particular those 
of Refs. \cite{Lu2009,Liu2010,Yi2013,sekiba2009}.
Concretely, we here show the evolution of the spectral function  $ A ({\vec{k},\omega})$ 
as a function of momentum $\vec{k}$, at low temperature: T = 20 K, and at half-filling: $n = 2$ (i.e. for the parent compounds).                              

In particular, in Fig. \ref{camino-GX}(a) notice that a relatively flat multiple-peak structure near $\Gamma$, evolves into a sharp peak near the 
Fermi level at $X$. In accordance with ARPES experiments along $\Gamma-X$, e.g. in LaOFe(P,As),(Fig. 4 of Ref.\cite{Lu2009}) in K$_{x}$Fe$_{2-y}$Se$_{2}$( Fig. 2 of Ref.\cite{Yi2013})
and BaFe$_{1.7}$Co$_{0.3}$As$_{2}$ compounds,( Fig. 2 of Ref.\cite{sekiba2009}) during this evolution, the main peak describes an S-like trajectory including an 
intermediate regime with a two peak-like structure, though in experiments the width and intensity of these features 
are material-dependent.                       

Along $\Gamma$ - $M$, in Fig. \ref{camino-GX}(b) notice how the broad relatively flat structure near $\Gamma$ and near $M$, for intermediate
BZ points develops a peak-like structure which describes a trajectory similar to that reported in ARPES experiments:( concretely, Fig. 1 of Ref.\cite{sekiba2009} and Fig. 1 of Ref.\cite{yang2010}) in particular, 
it resembles the evolution of the ARPES peak in LaFeAsO assigned to two bulk-related energy bands in  
Ref. \cite{yang2010}, where a second peak-like structure  closer to the Fermi level but related to surface-bands is also shown. Very recently, ARPES in FeSe single crystals reported \cite{nakayama2014} very similar energy distribution curves along  $\Gamma$ - $M$ in their Figures 1 and 2.

Notice that the spectral density results shown in Figs. \ref{camino-GX}(a) and \ref{camino-GX}(b)  include renormalization effects  due to the correlations which are taken into account by our analytical approach, concretely by the electron self-energies obtained, discussed in Supplementary Appendix B, which depend explicitly on crystal momentum $k$, as well as temperature and doping.  Band renormalization  effects such as these,  implying   changes  to the  non-interacting electron band structure of the minimal model proposed in Ref. \cite{raghu} (in particular, to their Fig. 2(a)), have also been  reported  in several  ARPES experiments:  
e.g. in  Refs. \cite{ding2011, Terashima, sekiba2009}.

\subsection{Effect of doping on the electronic structure}    
 
   \label{doping-effect}

 We will now refer to the effects of doping on the total density of states, and in particular the degree of accuracy presented by the results obtained with our approach, taking advantage of the experimental data available for comparison, such as Refs.\cite{neupane2011,tesisThiru}.  In Supplementary Appendix C we complement these results, showing also the 
 effects of doping on the spectral density function at the $\Gamma $ and $M$  symmetry points. 
 We carried out a systematic study covering a broad range of hole and electron doping values, 
   which respectively correspond to less or more electrons in the system than present at half-filling ($n=2$), which represents the parent compound. 
   In our calculations we fixed the band filling $n$, while the chemical potential $\mu$ was obtained self-consistently according to Eq. 17.

\begin{figure}[h!]
  \begin{center}
  \includegraphics[width=7cm]{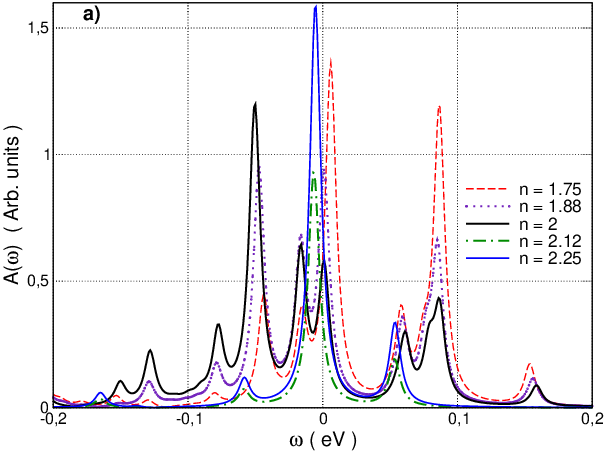}
 \includegraphics[width=7cm]{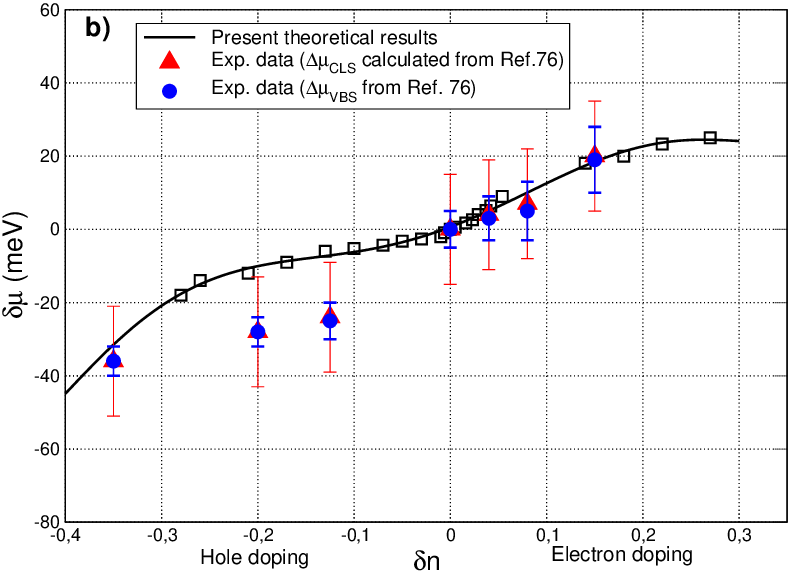}
    \caption[]{ Electron vs. hole doping. a) Effect on the total DOS. b) Chemical potential shift as a function of carrier (dopant) density,  
    with respect to the parent compound. 
     For the exp. values shown, obtained from reported\cite{neupane2011} core-level shift data $\Delta \mu_{CLS}$ (solid triangles) and valence-band shifts $\Delta \mu_{VBS}$ (solid circles), the carrier density is given as $x/2$ (per Fe):  either for BaFe$_{2-x}$Co$_{x}$As$_{2}$ or Ba$_{1-x}$K$_{x}$Fe$_{2}$As$_{2}$, 
     respectively for electron or hole doping.  
    At T = 20 K,    $U = V = 3.5 eV$.    Other parameters as in Fig.~\ref{dep-U}. }
  \label{electron-doping}
  \end{center}
  \end{figure}

   In Fig.\ref{electron-doping} (a)  we exhibit a comparison of the effect of hole vs. electron doping 
   on the  renormalized total density of states. 
    In agreement with ARPES experiments in Ref.\cite{neupane2011} (in particular, their Fig. 2(e)), we find that upon increasing hole-doping  a rigid-band-like shift of all peaks
     towards the Fermi level ensues, thus increasing spectral weight there ( for more details, see Suppl.Appendix C. In particular, in Figs. C.5 and C.6 the spectral density  results at $M$ are shown, and their comparison with experiments is discussed in detail).
   Meanwhile,   when the system is doped with electrons, $n=2.12$ and $n=2.25$ in the figure,  the bottom of the electronic structure is seen 
   to be shifted to lower binding energies.   Notice that for higher electron-doping values, e.g. $n= 2.25$,  a sharp intense peak appears at the Fermi level. 
   
  For comparison with the results of Figure 3 of Ref.\cite{neupane2011},
   in our  Fig \ref{electron-doping} (b) we plot the chemical potential shift: $\delta\mu = \mu(n) − \mu(n = 2)$ 
  with respect to the  undoped compound obtained in our approach as a function of the carrier (dopant)  density:  $\delta n = n - 2$. 
  As indicated in Ref.\cite{neupane2011}, experimental chemical potential shifts, which we denote $  \, \Delta \mu_{CLS} $,  can be obtained  from  the doping dependence 
  of the measured   As-$3d$  core level shifts ( $\Delta E_{CLS}$ ) reported in their Fig.3.b:   
by  using the formula   $  \, \Delta \mu_{CLS} \,  =  \, -  \Delta E_{CLS} \,  +   \, (  \Delta V_{M} +    \Delta E_{R} )   $        
, where   $ (  \Delta V_{M} +    \Delta E_{R}  )$    accounts for doping-related changes  in the Madelung potential  and in the core-carrier screening, respectively,  assuming
 that changes of As-valency upon doping  are  negligible. Using the  As-$3d$ core level shifts   and  the differences   $ (  \Delta V_{M} +    \Delta E_{R}  )$    reported in  Fig.3(b) 
 of Ref.\cite{neupane2011}, we calculated the experimental chemical potential shifts derived from core-level shift data  $  \, \Delta \mu_{CLS} \, $
  corresponding to the seven  compounds  experimentally studied, and have plotted  them in  our Figure \ref{electron-doping}(b).  
  A comparison with the theoretical results of our approach, 
shows that the asymmetry and especially the different signs exhibited by the chemical potential shifts for hole and electron doping are described,
   in agreement with experiments.\cite{coldea2008, neupane2011}  Notice that our theoretical results  also describe the correct monotonous dependence 
   of the chemical potential shift  as a function of doping for all samples, and  quantitative agreement is also obtained  within the  experimental  error bars of  $  \, \Delta \mu_{CLS} $ 
 for all the electron-doped samples  as well as for the sample   with  the  highest  hole-doping.  For completeness, in Figure \ref{electron-doping}(b) we also include the 
 experimental chemical shifts, denoted by  $ \, \Delta \mu_{VBS} \, $ , 
obtained  as $  \, \Delta \mu_{VBS} \, =  \, - \Delta E_{VBS} \, $   from the valence-band shift data $  ( \, \Delta E_{VBS} \, )  $ reported in  Fig.3(c) 
 of Ref.\cite{neupane2011}, which are consistent with  $  \, \Delta \mu_{CLS} $  obtained  from  the measured   As-$3d$  core level shifts.  
   The reported electron-hole asymmetry in  Ba-122 pnictides,\cite{neupane2011}  in particular in  BaFe$_{2}$As$_{2}$, 
   has been also observed in the superconducting phase.\cite{neupane2011,ding2008,ding2011, terashima2009}
 We checked that the chemical potential shift is almost independent of $V$,  consistently with our DOS results discussed in Section \ref{UV-effect}.

   Let us first stress one result: notice that  in spite of the band renormalization effects due to correlations  known to be  included in our level of approximation ( as shown in our self-energy analysis of Suppl.Appendix B),  we  obtain  electron-hole  asymmetric effects   such as experimentally observed\cite{neupane2011,tesisThiru}  
 i.e.  the bottom of the band  is shifted to lower, or higher, binding energies with respect to the Fermi level upon, respectively, 
   electron or hole doping,  as one would expect if the electronic structure is not changed by doping.     
  
  Next, we would like to stress that, 
   even though one might expect the values of the chemical potential shifts upon doping to be material-dependent, 
   we find that our    results in Fig.\ref{electron-doping}(b)  agree better with the reported data on Ba-122 compounds (Figure 3 of Ref. \cite{neupane2011}) 
   than the LDA calculations mentioned in Ref.\cite{neupane2011}: which required dividing their calculated  chemical potential shift values  
   by a factor of 4, in order to achieve quantitative agreement with experiments. 
   Previously, theoretical estimations\cite{cappelluti-2009}  of an LDA band structure renormalization, amounting to a factor of 2 division, 
   had been indicated as necessary to describe  with LDA the observed energy shifts of the electron and hole bands.\cite{coldea2008} 
  
  Thus, our analytical treatment for the effective two-band model  yields electronic structure results 
  describing  not only  the observed  opposite chemical potential shifts  for electron and  hole-doping,  
      but the chemical potential shift values  we obtain  are in quantitative agreement with the experimental results of Fig. 3 of  Ref.\cite{neupane2011}, respectively 
       for all electron doped samples in BaFe$_{2-x}$Co$_{x}$As$_{2}$ and for the sample with highest hole doping in Ba$_{1-x}$K$_{x}$Fe$_{2}$As$_{2}$ .
     
As mentioned in the Introduction, Lifshitz transitions have been reported for ferropnictides, in particular as a function of 
doping.\cite{cui2012,chang2011,ding2011} We have checked that the correlated two-orbital model indeed undergoes 
changes of topology of the Fermi surface upon doping increase, as exhibited in Figure C.8 of Suppl. Appendix C.
A detailed study of this topic is out of the scope of the present work, though we plan to address it in the future.

\subsection{Effects of temperature on the spectral properties}  
  
       \label{T-effect}     

Our analytical approach is specifically well adapted to shed light on the temperature dependence of the normal state spectral properties throughout 
the Brillouin zone, subject largely unstudied in ferropnictides, on which we were able to find interesting predictions. 

     \begin{figure}[h]
  \begin{center}
  \includegraphics[width=8.0cm]{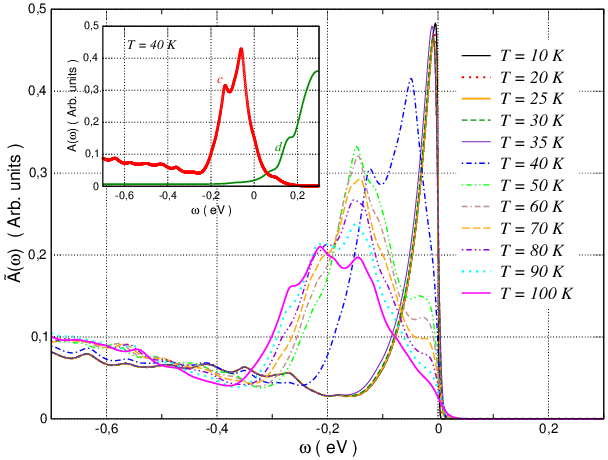}
    \caption[]{ Temperature dependence of $  \tilde{A} (\omega)$.
   Parameters: U = V = 3.5 eV , $ n = 2$.  Other parameters as in Fig.~\ref{dep-U}. 
   Inset: partial c- and d-band densities of states at 40 K.}
  \label{A-T}
  \end{center} 
  \end{figure}

               
               First, we will focus on the effect of the temperature on the total density of states. 
               We  found only low temperature density of states ARPES data  reported  for the parent compounds, e.g in Refs.\cite{Lu2009,Maletz2014}, 
               to compare  the results of this section,  while at room temperature 
                we only found a hard X-ray photoemission DOS result reported for undoped Ba-122.\cite{Jong2009}   
                 To allow a direct comparison with experimental data, in this section 
               all density of states results exhibited, denoted $\tilde{A} (\omega)$, have been convoluted with an energy resolution 
               of 30 meV and multiplied by $ f_{FD} (\omega) $, as in experiments.\cite{Lu2009, Jong2009, Maletz2014}.

               In Figure \ref{A-T} we show the total DOS we obtain for several temperatures varying from 10 K to 100 K. 
               As an inset, we plotted the corresponding partial densities of states at $T = 40 K$, 
               corresponding to the  two effective TB bands\cite{raghu} used in our model,  
                denoted by $c$ and $d$ and defined in Eq. \ref{partialdos}.

     In  Fig. \ref{A-T},  notice that with our simplified correlated model for pnictides the total DOS near the Fermi level 
     consists of a sharp intense peak, which the inset and our self-energy results presented in Supplementary Appendix B.1  
     show is mostly due to the renormalized c-effective band,  
     i.e. obtained in our approach by renormalization of the non-interacting c-band from  Eq. \ref{effectivebands} proposed by Raghu et al.\cite{raghu}, 
     which is the lowest energy effective band extending from -8 to 4 eV. 
      Our result agrees with previous work, which indicates that the peak originates mainly from Fe-3d states, 
      in particular:  ARPES experiments at low temperature in LaOFeP and LaOFeAs compounds (Figure 1 of Ref.\cite{Lu2009}) at $T = 20$ K interpreted with  DFT+LDA calculations, 
          ARPES at  $T = 28.7$K in  FeSe$_{1−x}$ (Figure 2 of Ref.\cite{Maletz2014}), 
         and hard X-ray photoemission experiments in BaFe$_{2}$As$_{2}$ (Figure 2 of Ref.\cite{Jong2009}) at 20K and 300K.   
        The low temperature results agree also with LDA+DFMT calculations in SmO$_{1-x}$F$_{x}$FeAs,\cite{craco2008}
     LDA+DFT calculations in LaFeAsO$_{1-x}$F$_{x}$,\cite{singh2008} BaFe$_{2}$As$_{2}$\cite{singh2008-2} and LiFeAs.\cite{singh2008-2}
     
     Regarding the evolution with temperature shown in  Fig. \ref{A-T},  
    our calculations predict that this intense sharp peak near the Fermi level should be clearly observed for temperatures $T<40$K. 
    By increasing temperature, we find that the peak near the Fermi level is shifted towards lower binding energy, 
    diminishes intensity and becomes broader, eventually developing additional structure. This agrees with the  observations    
     in BaFe$_{2}$As$_{2}$ (Figure 2 of \cite{Jong2009}) where, from the sharp peak near the Fermi level at 20 K, at room-temperature 
     a broad hump located at lower binding energies was reported. 

Notice in Fig. \ref{A-T}  that we also find new temperature dependent peaks in the DOS due to correlation effects, 
as our self-energy analysis in Supplementary Appendix B confirms.
Concretely,  if one  compares  the density of states  curves  corresponding  to  various temperatures in Fig. \ref{A-T}: 
 at 10 K the DOS is dominated by a single peak near the Fermi level,  at 40 K we obtain  two prominent peaks, shifted further below the Fermi level  w.r. the T= 10K result,  
 and  three  sizeable peaks  of  the density of states ( shifted still further below the Fermi level) are 
present at T= 100 K.    Therefore,  we are predicting a non-trivial temperature dependence of the total density of states due to correlations, 
which it would be interesting to investigate experimentally. 
      
     Furthermore, in connection with the location of the peak near the Fermi level, which may be material-dependent:  
     in Fig. \ref{A-T} we find it centered at $\omega \sim -10$ meV at  30 K, 
     which is similar to the peak position estimated from fits in Ref. \cite{Maletz2014} in FeSe$_{1−x}$, 
     placing it  at $\omega = -15$ meV at 30 K, in particular their Figure 2(d).


 Next, we will analize the evolution with temperature of the spectral density, analyzing the changes   in 
 $\tilde{A}(\vec{k}, \omega)$. In first term, we will  fix $\vec{k}$ at the relevant high-symmetry points of the BZ 
 which have been studied in ARPES experiments.\cite{liu2009, yang2010}

   \begin{figure}[h!]
  \begin{center}
  \includegraphics[width=6.5cm]{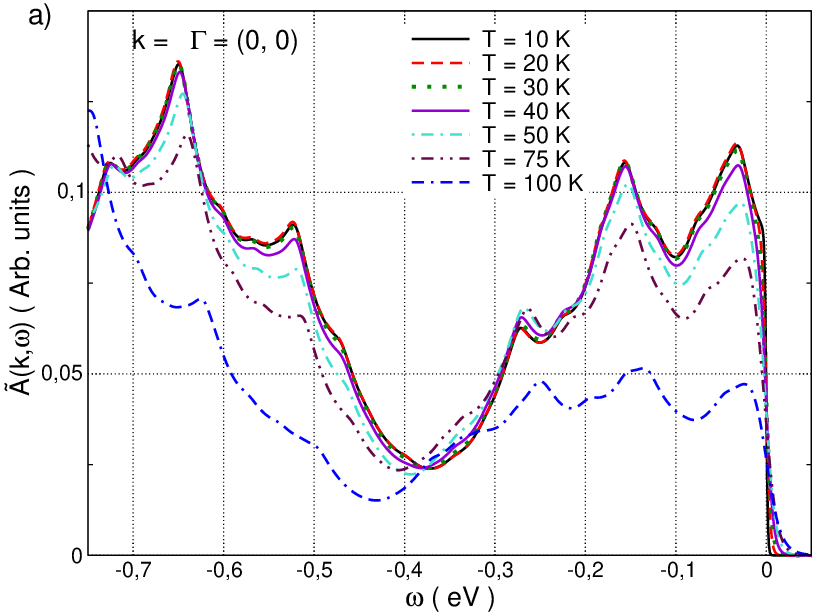}
  \includegraphics[width=6.5cm]{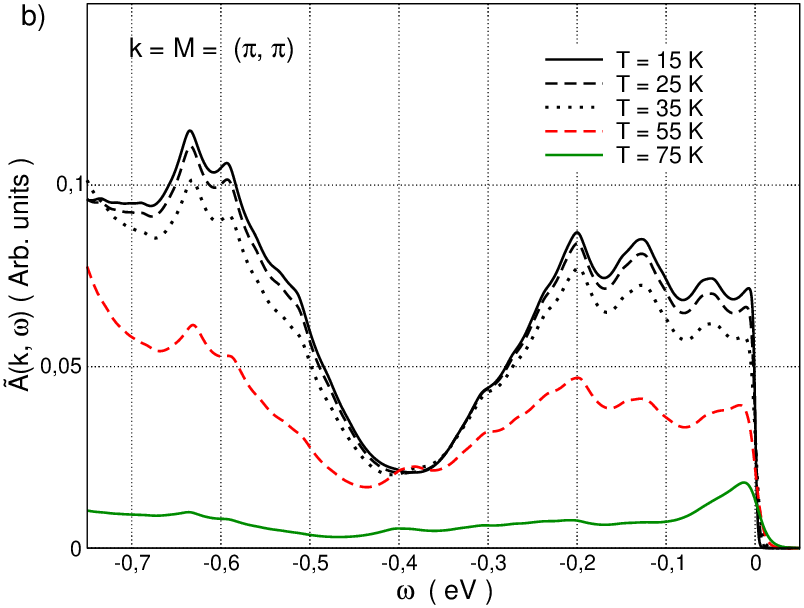}
    \caption[]{ a) Temperature dependence of $ \tilde{A} (k, \omega)$, at $\Gamma$; b) Temperature dependence of $ \tilde{A} (k, \omega)$, at $M$.  
    Same energy range used  for the ARPES data of Ref. \cite{yang2010}. 
    Parameters: U = V = 3.5 , $ n = 2$.  $\nu= 9$. Other parameters as in Fig.~\ref{dep-U}.}
  \label{Gamma}
  \end{center} 
  \end{figure}
 
 In Figure \ref{Gamma}(a), we show the evolution of the electronic structure with temperature at $\Gamma$.
 Notice that
 a number of prominent peaks emerge as temperature decreases. This is qualitatively what was observed in ARPES experiments 
 in LaFeAsO (Figure 7(a) of Ref. \cite{yang2010}) and CeFeAsO (Figure 4 of Ref. \cite{Liu2010}) compounds, where three peaks were also reported in  the  
 same energy range, though our results differ in the relative intensities.  
  Our results also agree qualitatively with recent ARPES experiments at $\Gamma$ closer to the Fermi level in
 NaFe$_{0.95}$Co$_{0.05}$As (Figure 3(a) of Ref.\cite{thiru2012}) and Ca$_{1-x}$Na$_{x}$Fe$_{2}$As$_{2}$ (Figure 2 of Ref. \cite{Evtushinsky2013}) compounds,
 who focus on the spectral density reduction at the Fermi level upon temperature increase.

  \begin{figure}[t!]
  \begin{center}
   \includegraphics[width=7cm]{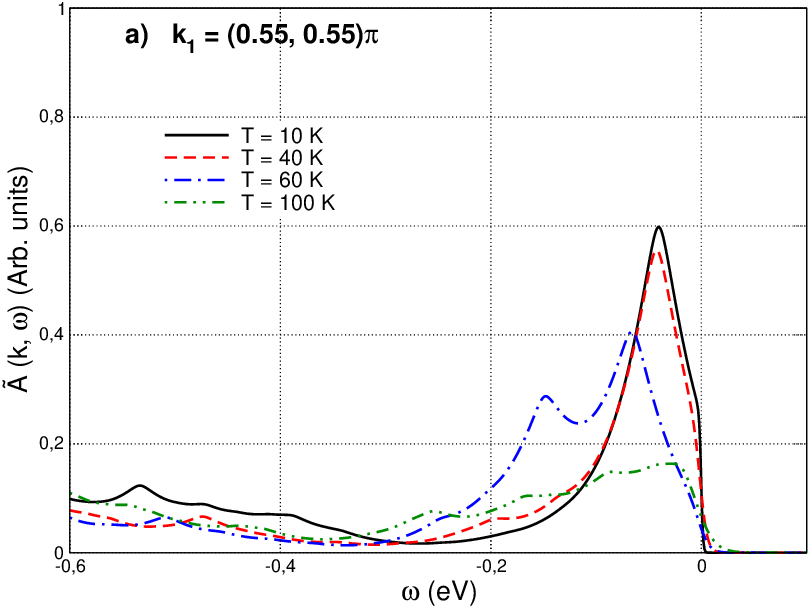}
  \includegraphics[width=7cm]{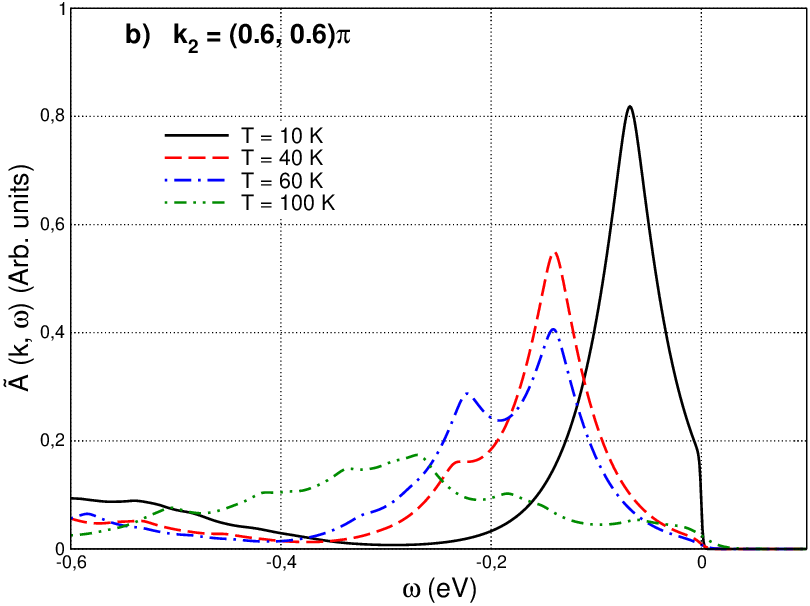}
  \includegraphics[width=7cm]{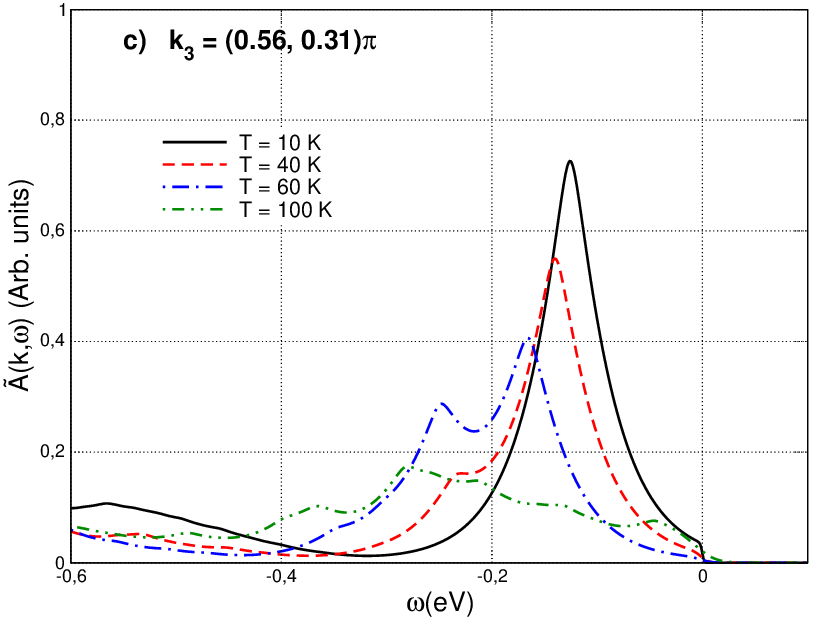}
     \caption[]{Temperature dependence of $ \tilde{A} (k, \omega)$, at the BZ points respectively indicated in each plot. 
     Parameters:  U = V = 3.5 eV , $n_{c} = 1.45$ and $n_{d} = 0.55$ ( $ n = 2 $), $\nu= 9$.  Other parameters as in Fig.~\ref{dep-U}.}
  \label{figure12}
  \end{center}
  \end{figure}  

 In Figure \ref{Gamma}(b)  we show the temperature evolution of the spectral density at M, to be compared with the reported ARPES data in  
 Figure 3 of Ref.\cite{Liu2010} and Figure 4(b) of Ref.\cite{zhou2010}, who focus on the energy range $[-0.2,0]$eV. 
 The description  of the low temperature ARPES data at M  with the effective two-orbital model is not as good as for $\Gamma$.  
 Though  the spectra at $M$ in the energy range $[-0.1,0]$eV
 exhibit a two-peak structure at low temperature which evolves to a broad single peak at higher temperature,  
 in agreement with ARPES experiments in CeFeAsO (Figure 3 of Ref.\cite{Liu2010}) and EuFe$_{2}$As$_{2}$ (Figure 4(b) of Ref.\cite{zhou2010}) compounds, 
 we find differences (e.g. a number of additional peaks at low temperature below -0.1 eV) 
 if we compare our results with experiments in a wider energy range.   
 The temperature evolution of the electronic structure at $\Gamma$ exhibits less relevant changes than at $M$, in agreement with experiments.
 
         In any case, the spectral function results at the high symmetry BZ points
  probed by ARPES, which are presented in Figs. \ref{Gamma}(a) and \ref{Gamma}(b), mainly exhibit the usual effects of thermal broadening, 
in correspondence  with the negligible  temperature-dependence of the self-energy  at these points, discussed in Supplementary Appendix B. 

Nevertheless, as discussed  in connection with Fig. \ref{A-T},  
  our model predicts a non-trivial temperature-dependence of the total density
  of states, to which the whole BZ contributes. To understand
  this fact,  we explored  throughout the whole BZ  the dependence on temperature of the k-dependent self-energy obtained in our approach, 
  and in Supplementary Appendix B exhibit our results.  
  Interestingly, for the parent compounds ($ n = 2 $) we have been able to  identify a number of  specific BZ points, not yet probed by ARPES,   
which exhibit  sizeable and non-trivial temperature dependent renormalization effects,  
allowing to explain  the temperature  dependence  predicted for the total DOS. 
  In particular, in Figure \ref{figure12}  we show the evolution of the spectral density function 
 with temperature  predicted at three specific BZ points ( $\vec{k_1}=(0.55\pi,0.55\pi)$, $\vec{k_2}=(0.6\pi,0.6\pi)$ and $\vec{k_3}=(0.56\pi,0.31\pi)  $), 
which we could identify as exhibiting the largest non-trivial temperature dependent renormalization effects due to correlations (for details, 
see Supplementary Appendix B). Notice also that some non-Fermi liquid features are visible at specific BZ points at higher temperatures: in fact,  
Figure \ref{figure12}(b) shows that  at $T=100 K$
 the spectral density function at $\vec{k_2}$  is mostly  dominated by the incoherent contribution, 
 a strong reduction of the quasiparticle peak having taken place.
  Our prediction could be tested by temperature dependent ARPES experiments, 
 and these effects should also be displayed by correlated multi-orbital models with more effective bands (though perhaps at larger correlation values). 
 In Supplementary Appendix C.1, we discuss how doping affects the results discussed above for $ n = 2$, and in particular how 
  it reduces the non-trivial temperature dependent renormalization effects by correlations in the density of states, with differences between electron and hole doping. 
 We also show how doping affects the k-dependence discussed for the parent compounds  in connection with Fig. \ref{figure12}. Our results indicate that a 
 a subtle interplay between correlations, temperature and doping  gives rise to the non-trivial temperature dependence effects in the spectral properties.

\begin{figure}[h]
  \begin{center}
   \includegraphics[width=7cm]{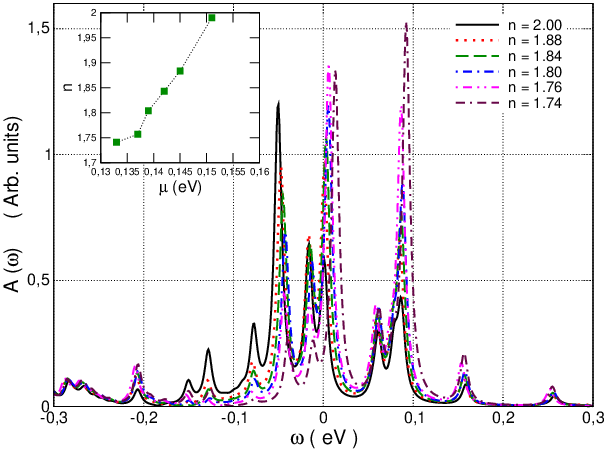}
       \caption[]{Quasiparticle weights at the Fermi level, $Z(\vec{k})$, shown at a set BZ points (located as indicated in the inset), for temperatures: $T = 40, 60, 100$ K. 
       $U = V = 3.5$ eV and other parameters as in Fig.\ref{dep-U}. $k_{0}=(\pi/2, \pi/2)$, $k_{1}=(0.55\pi, 0.55\pi)$, $k_{2}=(0.6\pi, 0.6\pi)$, $k_{3}=(0.56\pi, 0.31\pi)$,$k_{4}=(\pi, 0.5\pi)$ and $k_{5}=(0.5\pi,0)$.}
  \label{weight}
  \end{center}
  \end{figure}

Finally, the quasiparticle weights Z(k) at the Fermi level, defined as $Z^{-1}(\vec{k})=1-\frac{\partial}{\partial\omega}Re\Sigma_{c}(\vec{k},\omega)\vert_{\omega=0}$, 
are shown at different BZ points  in Fig.\ref{weight}, exhibiting the non-trivial effect of temperature along the BZ: clearly larger at some points like   
 $k_{3}=(0.56\pi, 0.31\pi)$, which would be interesting to probe by ARPES. 
Note that the mass renormalization ($\frac{m}{m^{*}} \varpropto Z$) 
obtained in our approach corresponds to a correlated metal, while the range of quasiparticle weights obtained along the BZ agrees  
with LDA+DMFT results\cite{shorikov2009} who estimated a quasiparticle weight renormalization factor $Z \sim 0.8$ for $J = 0$, 
as well as with T=0 slave-spin mean-field studies of two- and four-orbital models.\cite{qsi2}

\section{Conclusions}\label{conclusions}

Though a large amount of experimental and theoretical research work has been done since 2008, the true nature of electron correlations in ferropnictides 
remains an open question.  Here, using a simplified model based on two correlated effective bands and 
an analytical approximation to decouple the equations of motion for the electron Green's functions,
 we determined the normal state spectral density function and the total density of states.
With model parameters  in the range relevant for iron-based superconductors, we could describe:
(i) the momentum dependence measured in ARPES along $\Gamma-X$, and the main features along $\Gamma-M$;
(ii) an asymmetric effect of electron and hole doping, with the characteristics observed in ARPES experiments. In particular, 
the band structure renormalization due to the correlation effects captured by our analytical approach, 
allows to obtain a direct agreement of our calculations with the experimental chemical potential shifts reported, 
with  the correct  order of magnitude  in contrast to  LDA calculations. 
(iii) We found that an important  spectral weight reduction near the Fermi level takes place if temperature is increased,  and 
the evolution with temperature of the total density of states which we predict is consistent with the relatively 
few available data  from ARPES and hard X-ray photoemission experiments.  
(iv) The conduction electron mass renormalization obtained corresponds to a correlated metal, 
while the range of quasiparticle weights at the Fermi level which we obtain along the Brillouin zone  
agrees with theoretical LDA+DMFT results which estimated $Z \sim 0.8$ in the absence of Hund coupling.

Among the results obtained, we want to stress an interesting prediction which might be tested experimentally: our work not only predicts  
a non-trivial temperature dependence of the total density of states, due to correlation effects. We could trace the origin of this behaviour 
to specific Brillouin zone regions where the effect of temperature on the renormalization is amplified,     
by exploring the k-dependent self-energy obtained in our approach throughout the Brillouin zone.
It would be interesting to have this prediction tested, by temperature dependent ARPES experiments in the specific BZ points 
we explicitly identified for undoped compounds.
  
Our calculated Green's functions might be also employed to evaluate  other normal state physical properties of iron pnictides, at our level of approximation: 
 which has the advantage of enabling to describe temperature and k-dependent results,  
 in contrast to other theoretical techniques
 where the self-energy is local.


We have shown that our results describe well most of the main common features of the bulk electronic structure observed, 
not only 1111 and 122 ferropnictides, but also in FeSe compounds which we believe is related to the fact that 
the two correlated effective bands of the model capture the relevant details of the 3d-Fe orbitals near the Fermi level,  
lying the As or Se orbitals further below in energy.      

The simplified pnictides model and analytical approach used in the present work, 
are suitable for extensions to describe details of magnetotransport in FeSe compounds, as
well as adapting them to describe electronic properties of BiS$_{2}$-based layered compounds.
Our approach could also be useful to address the nature of correlations in Hund metals and its relationship with Mott-physics
 in other multi-orbital systems \cite{fanfarillo2015}, problems we plan to study in the future.

\section{Acknowledgments}

R.C. acknowledges I. Eremin for useful discussions on the manuscript. C.I.V. is Investigadora Cient{\'{\i}}fica of CONICET (Argentina). J.D.Q.F. has a fellowship from CONICET. C.I.V. acknowledges support from CONICET (PIP 0702) and ANPCyT (PICT'38357; PICT Redes'1776), and the hospitality of ICTP (International Centre  for Theoretical Physics, Trieste)  and Dipto. di Fisica, Universit\'a di Salerno, where part of this collaboration work was done. J.D.Q.F. acknowledges the hospitality of IIP (International Institute of Physics, Natal).


\vspace{2cm}

\appendix 

\section{Analytical approximation: decoupling of the Green's function equations of motion.}\label{apen1}

As mentioned in section \ref{calculations}, for the model Hamiltonian given by Eqs. \ref{Hamiltonian}, \ref{Hamiltonian0} and \ref{Vint} 
we calculated the equations of motion (EOM) of the Green's functions $G(\vec{k},\omega)$ and $F(\vec{k},\omega)$, respectively corresponding to c- and d-electrons defined in  Eqs. \ref{g-retarded} and \ref{f-retarded}, using Zubarev's formalism\cite{zubarev}.   We obtained the 
set of exact equations of motion detailed in Eqs. \ref{eqG} and \ref{eqF}. 
Furthermore, the equations of motion for the three $\Gamma_{i}$ (i=1,3) functions coupled to $G(\vec{k},\omega)$, 
and for $\Gamma_{i}$ (i=4,6) coupled to $F(\vec{k},\omega)$, introduce coupling to new higher-order Green's functions in the problem. 
In this appendix we will detail the RPA-like analytical approximation used to obtain the Green's functions $G(\vec{k},\omega)$ and $F(\vec{k},\omega)$, required 
to calculate the spectral density and total density of states defined in Eqs. \ref{partial} and \ref{totalA}.

\subsection{First order solution of EOM: Hartree-Fock decoupling.}

 First, we will explain how one could close and solve the coupled set of equations of motion in lowest order, which would correspond to 
the Hartree-Fock approximation. 

The approximation to be used in Eqs. \ref{eqG} and \ref{eqF} for  the $\Gamma_i $ ( i = 1,6) Green's functions which appear coupled to $ G(\vec{k},\omega)$ and  $ F(\vec{k},\omega)$, respectively, 
 basically consists  in approximating  $\Gamma_i $ ( i = 1,6) as  proportional to  $ G(\vec{k},\omega)$ and  $ F(\vec{k},\omega)$,  with mean values of the occupation numbers of the c- and d-bands  appearing as proportionality factors, thus obtaining 
a closed system of equations of motion.  Below, we show the approximation explicitly  for  ${\Gamma}_{1}(k_{1},k_{2},k,\omega)$:

\begin{eqnarray}\label{hf1}
& {\Gamma}_{1}(k_{1},k_{2},k,\omega)  \equiv  {\Gamma}^{ccc}_{\sigma, \overline{\sigma},\overline{\sigma}}(k_{1},k_{2},k,\omega) =  \nonumber \\ 
& =  \ll c_{k_{2},\sigma} c^{\dagger}_{k_{1}, \overline{\sigma}}  c_{k_{1}-k_{2}+k, \overline{\sigma}}  ;  c^{\dagger}_{k \sigma} \gg (\omega)\nonumber \\
& \sim    < c^{\dagger}_{k_{1}, \overline{\sigma}}  c_{k_{1}-k_{2}+k, \overline{\sigma}}  > \ll c_{k_{2}, \sigma} ;  c^{\dagger}_{k \sigma} \gg(\omega)
 \nonumber \\
& \sim    < c^{\dagger}_{k_{1}, \overline{\sigma}} c_{k_{1}-k_{2}+k, \overline{\sigma}} > \delta_{k_{1}, k_{1}-k_{2}+k } \,  
\ll c_{k_{2}, \sigma} ;  c^{\dagger}_{k \sigma} \gg (\omega)
  \nonumber \\
& \sim    < n_{k_{1},\overline{\sigma}} > \delta_{k, k_{2}} \, G_{\sigma}(k,\omega)    
\end{eqnarray}

where: $ n_{k_{1},\overline{\sigma}} = < c^{\dagger}_{k_{1}, \overline{\sigma}} c_{k_{1}, \overline{\sigma}} >$.

\noindent Similarly, in this first level of approximation:

\begin{eqnarray}\label{hfrest}
& {\Gamma}_{2}(k_{1},k_{2},k,\omega)  \equiv  {\Gamma}^{cdd}_{\sigma, \sigma ,\sigma}(k_{1},k_{2},k,\omega) \nonumber \\
& \sim    < N_{k_{1},\sigma} > \delta_{k, k_{2}} \, _{\sigma}(k,\omega) \nonumber \\
& {\Gamma}_{3}(k_{1},k_{2},k,\omega) \equiv  {\Gamma}^{cdd}_{\sigma, \overline{\sigma} ,\overline{\sigma}}(k_{1},k_{2},k,\omega) \nonumber \\
& \sim    < N_{k_{1},\overline{\sigma}} > \delta_{k, k_{2}} \, G_{\sigma}(k,\omega) 
\end{eqnarray}

where: $ N_{k_{1},\sigma} = < d^{\dagger}_{k_{1}, \sigma} d_{k_{1}, \sigma} >$.

\noindent Therefore, replacing this approximation for $\Gamma_i $ ( i = 1,3) in  Eq. \ref{eqG} one has:

\begin{align}
[\omega - E_{c}(k)] G_{\sigma}(k, \omega)   =   \frac{1}{2\pi} + \sum_{k_{1}, k_{2}} {\frac{U}{N}\delta_{k_{1},k_{2}}} \langle {n}_{{k}_{1},\overline{\sigma}}\rangle G_{\sigma}(k, \omega)  \nonumber \\ 
+ \frac{V}{N}\delta_{k_{2},k}\left(  \langle {N}_{{k}_{1},\overline{\sigma}}\rangle + \langle {N}_{{k}_{1},\sigma}\rangle \right) G_{\sigma}(k, \omega)
\raisetag{1pt}
\end{align}

\noindent from which one directly obtains: 

\begin{equation}
G^{H.F}_{\sigma}(k, \omega)  \sim \frac{1}{  2\pi\left[  \omega -  E_{c}(k)  - \sum_{k_{1}}{ \left(\frac{U}{N} \langle {n}_{{k}_{1},\overline{\sigma}}\rangle +  \frac{V}{N} \langle {N}_{{k}_{1}}\rangle \right) }   \right]  } 
\raisetag{2pt}
\end{equation}

\noindent where: $\langle {N}_{{k}_{1}}\rangle \equiv  \langle {N}_{{k}_{1}\uparrow}\rangle + \langle {N}_{{k}_{1}\downarrow}\rangle $.

On the other hand, using this approximation for $\Gamma_i $ ( i = 4,6) one has: 
\begin{eqnarray}\label{set3}
& {\Gamma}_{4}(k_{1},k_{2},k,\omega) \equiv  {\Gamma}^{ddd}_{\sigma, \overline{\sigma},\overline{\sigma}}(k_{1},k_{2},k,\omega) \nonumber \\
 & \sim    < N_{k_{1},\overline{\sigma}} > \delta_{k, k_{2}} \, F_{\sigma}(k,\omega) \nonumber \\
& {\Gamma}_{5}(k_{1},k_{2},k,\omega)  \equiv  {\Gamma}^{dcc}_{\sigma, \sigma ,\sigma}(k_{1},k_{2},k,\omega) \nonumber \\
& \sim    < n_{k_{1},\sigma} > \delta_{k, k_{2}} \, F_{\sigma}(k,\omega) \nonumber \\
& {\Gamma}_{6}(k_{1},k_{2},k,\omega)  \equiv  {\Gamma}^{dcc}_{\sigma, \overline{\sigma} ,\overline{\sigma}}(k_{1},k_{2},k,\omega) \nonumber \\ 
& \sim    < n_{k_{1},\overline{\sigma}} > \delta_{k, k_{2}} \, F_{\sigma}(k,\omega) \nonumber \\
\end{eqnarray} 

\noindent which, by replacement in Eq. \ref{eqF}, leads to: 

\begin{equation}
F^{H.F}_{\sigma}(k, \omega)  \sim \frac{1}{  2\pi\left[  \omega -  E_{d}(k)  - \sum_{k_{1}}{ \left(\frac{U}{N} \langle {N}_{{k}_{1},\overline{\sigma}}\rangle +  \frac{V}{N} \langle {n}_{{k}_{1}}\rangle \right) }   \right]  } 
\end{equation}

\noindent Finally, considering that: 

\begin{eqnarray}
 \frac{1}{N}\sum_{k_{1}}  \langle {n}_{{k}_{1},\overline{\sigma}}\rangle  & = &  \langle {n}_{i,\overline{\sigma}}\rangle \equiv  \langle {n}_{c,\overline{\sigma}}\rangle \, \,  \,  \forall i \\
 \frac{1}{N}\sum_{k_{1}}  \langle {N}_{{k}_{1}}\rangle  & =  &  \langle {N}_{i}\rangle \equiv \langle {n}_{d}\rangle \,  \,  \, \forall i
\end{eqnarray}

\noindent one can  rewrite the Green's functions in this first order (Hartree-Fock) order of approximation as:   

\begin{equation}
G^{H.F}_{\sigma}(k, \omega)  \cong \frac{1}{  2\pi\left[  \omega -  E_{c}(k)  - \left(U\langle {n}_{c,\overline{\sigma}}\rangle -  V\langle {n}_{d}\rangle \right)    \right]  } 
\end{equation}

\noindent and,

\begin{equation}
F^{H.F}_{\sigma}(k, \omega)  \cong \frac{1}{  2\pi\left[  \omega -  E_{d}(k)  - \left(U\langle {n}_{d,\overline{\sigma}}\rangle -  V\langle {n}_{c}\rangle \right)    \right]  } 
\end{equation}

\subsection{Second-order solution of EOM: RPA decoupling}

In the following, we will describe the second-order decoupling  and solution of the equations of  motion for 
 $G(\vec{k},\omega)$ and $F(\vec{k},\omega)$, which we used 
 to calculate all spectral density and total density of states results presented in Section \ref{results}.

As mentioned above, the exact equation of motion for  $G(\vec{k},\omega)$, Eq. \ref{eqG}, 
involves sums over two k-indices of the three coupled Green's functions:  $\Gamma_{i}$ (i=1,3) defined in Eqs.\ref{set1}. 
Now, instead of approximating these coupled Green's functions directly in Eq. \ref{eqG}, 
as in the first order (HF)  solution,  we determine the three equations of motion 
 for  the $\Gamma_{i}$ (i=1,3) Green's  functions, respectively.  Concretely, for
 $ {\Gamma}_{1}(k_{1},k_{2},k,\omega)  \,  =  \,  
  \ll~c_{k_{2},\sigma} c^{\dagger}_{k_{1}, \overline{\sigma}}  c_{k_{1}-k_{2}+k, \overline{\sigma}}  ;  c^{\dagger}_{k \sigma} \gg (\omega)$ 
 we obtain the following equation of  motion:
   
\begin{eqnarray}\label{EM-gamma1}
&        \omega   {\Gamma}_{1}(k_{1},k_{2},k,\omega) \,  =  \, \frac{ \delta_{k,k_{2}}}{2\pi} \langle n_{k_{1},\overline{\sigma}} \rangle  \, + \,\nonumber \\
&  \ll \left[ \Gamma_{1}(k_{1},k_{2},k), \mathcal{H} \right]  ;  c^{\dagger}_{k \sigma} \gg (\omega)
\end{eqnarray}

\noindent where  $\left[ \Gamma_{1}, \mathcal{H} \right]  =  \left[ \Gamma_{1}, \mathcal{H}_{0} \right] + \left[ \Gamma_{1}, V_{int} \right]   $ introduces coupling of the EOM to new higher order Green's functions.
In particular, the term  $\left[ \Gamma_{1}, \mathcal{H}_{0} \right]$  
couples new Green's functions $<< A; c^{\dagger}_{k \sigma} >> $  where $A$ involves three operator products: one creation and two annihilation ones,  e.g.:  $\Gamma^{cdd}_{\sigma, \overline{\sigma}, \overline{\sigma}}(k'_{1},k'_{2},k,\omega)   \equiv  \ll c^{\dagger}_{k'_{1},\overline{\sigma} }c_{k'_{1}-k_{2}+k, \overline{\sigma} }   c_{k'_{2},\sigma} ; c^{\dagger}_{k, \sigma}\gg (\omega)$.
On the other hand, the term $\left[ \Gamma_{1}, V_{int} \right] $ in Eq. \ref{EM-gamma1} couples four new higher order Green's functions, 
each involving five operator products in $A$: two creation and three annihilation  ones, such as  $c_{\sigma}c^{\dagger}_{\overline{\sigma}}c_{\overline{\sigma}}c^{\dagger}_{\overline{\sigma}}c_{\overline{\sigma}}$,
$c_{\sigma}c^{\dagger}_{{\sigma}}c_{{\sigma}}c^{\dagger}_{\overline{\sigma}}c_{\overline{\sigma}}$, \, \,    $c_{\sigma}c^{\dagger}_{\overline{\sigma}}c_{\overline{\sigma}}d^{\dagger}_{{\sigma}}d_{{\sigma}}$
and  $c_{\sigma}c^{\dagger}_{\overline{\sigma}}c_{\overline{\sigma}}d^{\dagger}_{\overline{{\sigma}}}d_{\overline{\sigma}}$. 
Similar results are obtained when calculating the equations of motion for  $ {\Gamma}_{2}(k_{1},k_{2},k,\omega)$  \, and  \,  $ {\Gamma}_{3}(k_{1},k_{2},k,\omega)$.   

Analogously,  the exact equation of motion for  $F(\vec{k},\omega)$, Eq. \ref{eqF}, 
involves sums over two k-indices of three coupled Green's functions:  $\Gamma_{i}$ (i= 4,6) defined in Eqs.\ref{set2},  
whose  own equations of motion indirectly couple  $F(\vec{k},\omega)$ to new higher-order Green's functions in the problem. 

To close the coupled set of equations of motion at this second-order level, we used the following approximation:
all new higher-order Green's functions introduced in each subset of equations of motion  for $\Gamma_{i}$ $(i = 1,3)$, were approximated  
in mean-field in terms of $\Gamma_{i}$ $(i = 1,3)$  and $G(\vec{k},\omega)$ (introducing appropriate average values), yielding a $4\times4$ closed set of equations of motion. 
We proceeded likewise for the subset of equations for  $\Gamma_{i}$ $(i = 4,6)$ related to  $F(\vec{k},\omega)$.

To exemplify the decoupling scheme adopted to close the set of EOM at this level, 
below we show the approximation used for the  higher-order Green's functions involving three operator products: $\Gamma^{cdd}_{\sigma, \overline{\sigma}, \overline{\sigma}}(k'_{1},k'_{2},k,\omega)  = \, \ll~c_{k'_{2}, \sigma }c^{\dagger}_{k'_{1}, \overline{\sigma}} c_{k'_{1}-k_{2}+k, \overline{\sigma}} \, ; \, c^{\dagger}_{k, \sigma} \gg (\omega)$, according
 to the definition in Eq.\ref{definition}. Approximating this Green's function in mean field   yields:
\begin{eqnarray*}
  \Gamma^{cdd}_{\sigma, \overline{\sigma}, \overline{\sigma}}(k'_{1},k'_{2},k,\omega) &   \simeq & \langle  c^{\dagger}_{k'_{1},\overline{\sigma} }c_{k'_{1}-k_{2}+k, \overline{\sigma} }  \rangle  \ll c_{k'_{2},\sigma} ; c^{\dagger}_{k, \sigma}\gg(\omega)\\
                                                                &  \simeq & \delta_{k'_{2}k} \langle n_{k'_{1}, \overline{\sigma}} \rangle  \ll c_{k'_{2},\sigma} ; c^{\dagger}_{k,\sigma}\gg(\omega)\\
                                                                &  \simeq & \langle n_{k'_{1},\overline{\sigma}} \rangle G_{\sigma} (k,\omega)                                       
 \end{eqnarray*}

 We now illustrate the decoupling of the  higher-order Green's functions involving five operator products, 
 by two examples: $\ll~c_{k'_{2}, \sigma }c^{\dagger}_{k'_{1}, \overline{\sigma}} c_{k_{2}-k'_{2}+k'_{1}, \overline{\sigma}} c^{\dagger}_{k_{1}, \overline{\sigma} } c_{k_{1}- k_{2}+k, \overline{\sigma}} \, ; \, c^{\dagger}_{k, \sigma}~\gg $
  and $\ll~c_{k'_{2}, \sigma }c^{\dagger}_{k'_{1}, \overline{\sigma}} c_{k_{1}-k_{2}+k, \overline{\sigma}} d^{\dagger}_{k'_{1}, {\sigma} } d_{k_{2}- k'_{2}+k'_{1}, {\sigma}} \, ; \, c^{\dagger}_{k, \sigma}~\gg $. Approximating these Green's functions in mean field    yields:
 \begin{gather*}
 \ll c_{k'_{2}, \sigma }c^{\dagger}_{k'_{1}, \overline{\sigma}} c_{k_{2}-k'_{2}+k'_{1}, \overline{\sigma}} c^{\dagger}_{k_{1}, \overline{\sigma} } c_{k_{1}- k_{2}+k, \overline{\sigma}} \, ; \, c^{\dagger}_{k, \sigma} \gg(\omega) \nonumber \\ 
\simeq \langle c^{\dagger}_{k'_{1}, \overline{\sigma}} c_{k_{2}-k'_{2}+k'_{1}, \overline{\sigma}} \rangle \ll c_{k'_{2}, \sigma }c^{\dagger}_{k_{1}, \overline{\sigma} } c_{k_{1}- k_{2}+k, \overline{\sigma}} \, ; \, c^{\dagger}_{k, \sigma} \gg(\omega)\nonumber\\
+ \langle c^{\dagger}_{k_{1}, \overline{\sigma}} c_{k_{1}-k_{2}+k, \overline{\sigma}} \rangle \ll c_{k'_{2}, \sigma }c^{\dagger}_{k_{1}, \overline{\sigma} } c_{k_{1}- k_{2}+k, \overline{\sigma}} \, ; \, c^{\dagger}_{k, \sigma} \gg (\omega)\nonumber\\
\simeq \delta_{k_{1}, k'_{2}} \langle n_{k'_{1},\overline{\sigma}} \rangle {\Gamma}^{ccc}_{\sigma, \overline{\sigma},\overline{\sigma}}(k_{1},k'_{2},k,\omega) \\
+ \delta_{k, k_{2}} \langle n_{k_{1},\overline{\sigma}} \rangle {\Gamma}^{ccc}_{\sigma, \overline{\sigma},\overline{\sigma}}(k'_{1},k'_{2},k,\omega)  \nonumber \\
\simeq \delta_{k_{1}, k'_{2}} \langle n_{k'_{1},\overline{\sigma}} \rangle {\Gamma}_{1}(k_{1},k'_{2},k,\omega) + \delta_{k, k_{2}} \langle n_{k_{1},\overline{\sigma}} \rangle {\Gamma}_{1}(k'_{1},k'_{2},k,\omega)  \nonumber
\end{gather*}

\noindent and, 

\begin{gather*}
 \ll c_{k'_{2}, \sigma }c^{\dagger}_{k'_{1}, \overline{\sigma}} c_{k_{1}-k_{2}+k, \overline{\sigma}} d^{\dagger}_{k'_{1}, {\sigma} } d_{k_{2}- k'_{2}+k'_{1}, {\sigma}} \, ; \, c^{\dagger}_{k, \sigma} \gg (\omega)\nonumber \\ 
\simeq \langle c^{\dagger}_{k_{1}, \overline{\sigma}} c_{k_{1}-k_{2}+k, \overline{\sigma}} \rangle \ll c_{k'_{2}, \sigma }d^{\dagger}_{k'_{1}, {\sigma} } d_{k_{2}- k'_{2}+k'_{1}, {\sigma}} \, ; \, c^{\dagger}_{k, \sigma} \gg (\omega)\nonumber\\
+\langle d^{\dagger}_{k'_{1}, {\sigma} } d_{k_{2}- k'_{2}+k'_{1}, {\sigma}} \rangle \ll c_{k'_{2}, \sigma }c^{\dagger}_{k_{1}, \overline{\sigma} } c_{k_{1}- k_{2}+k, \overline{\sigma}} \, ; \, c^{\dagger}_{k, \sigma} \gg(\omega) \nonumber\\
\simeq \delta_{k_{2}, k}\langle n_{k_{1},\overline{\sigma}} \rangle {\Gamma}_{2}(k'_{1},k'_{2},k,\omega) + \delta_{k_{2},k'_{2}}\langle N_{k'_{1}, {\sigma}} \rangle {\Gamma}_{1}(k_{1},k'_{2},k,\omega)
\end{gather*}

All other higher-order Green's functions were decoupled following the same procedure.    
We proceeded likewise for the subset of equations for  $\Gamma_{i}$ $(i = 4,6)$ related to  $F(\vec{k},\omega)$.

   Finally, the equations of motion for the six Green's functions $\Gamma_{i}$ $(i = 1,6)$  in RPA approximation read:

\begin{widetext}


Equation of motion for ${\Gamma}_{1}({k}_{1},{k}_{2},k,\omega)$: 

\begin{eqnarray}
\omega  {\Gamma}_{1}({k}_{1},{k}_{2},k,\omega) & \cong & \frac{{\delta}_{{k}_{2},k}}{2\pi}\langle {n}_{{k}_{1},\overline{\sigma}} \rangle +
              \left[ {E}_{c}({k}_{1}-{k}_{2}+k) - {E}_{c}({k}_{1}) + {E}_{c}({k}_{2}) + U({n}_{c} + 1) \right] {\Gamma}_{1}({k}_{1},{k}_{2},k,\omega) \nonumber \\
 & + & \{ 2V\langle {n}_{{k}_{1},\overline{\sigma}}\rangle{n}_{d} + (2+\langle {n}_{{k}_{1},\overline{\sigma}} \rangle)U{n}_{c}+ U(\langle {n}_{{k}_{1},\overline{\sigma}} \rangle - \langle {n}_{{k}_{1}-{k}_{2}+k,\overline{\sigma}} \rangle)\} {G}_{{\sigma}}(k, \omega) \nonumber \\
 & +  & \frac{V}{N} \left(\langle {n}_{{k}_{1},\overline{\sigma}} \rangle - \langle {n}_{{k}_{1}-{k}_{2}+k,\overline{\sigma}} \rangle \right) \left[\sum_{{k}^{'}_{1}}{{\Gamma}_{2}({k}^{'}_{1},{k}_{2},k,\omega)} + \sum_{{k}^{'}_{1}}{{\Gamma}_{3}({k}^{'}_{1},{k}_{2},k,\omega)}\right] \label{gamma1}
 \end{eqnarray}

 Equation of motion for ${\Gamma}_{2}({k}_{1},{k}_{2},k,\omega)$:

\begin{gather}
\omega  {\Gamma}_{2}({k}_{1},{k}_{2},k,\omega)  \cong   \frac{{\delta}_{{k}_{2},k}}{2\pi}\langle {N}_{{k}_{1},{\sigma}} \rangle + \left[ {E}_{d}({k}_{1}-{k}_{2}+k) - {E}_{d}({k}_{1}) + {E}_{c}({k}_{2}) + U{n}_{c} + 2{n}_{d}(U+V) - V \right] {\Gamma}_{2}({k}_{1},{k}_{2},k,\omega) \nonumber \\
                                          +       \{ (U-V)\langle {N}_{{k}_{1},{\sigma}}\rangle + 2V\langle {N}_{{k}_{1},{\sigma}} \rangle {n}_{d}- V(\langle {N}_{{k}_{1},{\sigma}} \rangle - \langle {N_{{k}_{1}-{k}_{2}+k,{\sigma}} \rangle) +V \} {G}_{{\sigma}}(k, \omega) } \nonumber \\
                                          +       \frac{V}{N} \left(  \langle {N}_{{k}_{1},{\sigma}} \rangle - \langle {N}_{{k}_{1}-{k}_{2}+k,{\sigma}}  \rangle \right) \left[\sum_{{k}^{'}_{1}} { \Gamma_{1}({k}^{'}_{1},{k}_{2},k,\omega)} \right] + \frac{U}{N} \left(\langle {N}_{{k}_{1},{\sigma}} \rangle - \langle {N}_{{k}_{1}-{k}_{2}+k,{\sigma}} \rangle \right) \left[\sum_{{k}^{'}_{1}}{{\Gamma}_{3}({k}^{'}_{1},{k}_{2},k,\omega)}\right] \label{gamma2} 
 \raisetag{1pt}
 \end{gather}

 Equation of motion for ${\Gamma}_{3}({k}_{1},{k}_{2},k,\omega)$: 
 
\begin{gather}
\omega  {\Gamma}_{3}({k}_{1},{k}_{2},k,\omega) \cong  \frac{{\delta}_{{k}_{2},k}}{2\pi}\langle {N}_{{k}_{1},\overline{\sigma}} \rangle + \left[ {E}_{d}({k}_{1}-{k}_{2}+k) - {E}_{d}({k}_{1}) + {E}_{c}({k}_{2}) + U{n}_{c} + 2V n_{d} + V \right] {\Gamma}_{3}({k}_{1},{k}_{2},k,\omega) \nonumber \\
 + \{ U\langle {N}_{{k}_{1},\overline{\sigma}}\rangle {n}_{c} + V(1 - \langle {n}_{{k}_{1},\overline{\sigma}} \rangle -  \langle {n}_{{k}_{1}-{k}_{2}+k,\overline{\sigma}} \rangle)n_{d} + V(\langle {n}_{{k}_{1},\overline{\sigma}} \rangle \} {G}_{{\sigma}}(k, \omega) \nonumber \\
 +  \frac{V}{N} \left(\langle {N}_{{k}_{1},\overline{\sigma}} \rangle - \langle {N}_{{k}_{1}-{k}_{2}+k,\overline{\sigma}} \rangle \right) \left[\sum_{{k}^{'}_{1}}{{\Gamma}_{1}({k}^{'}_{1},{k}_{2},k,\omega)} \right] +  \frac{U}{N} \left(\langle {N}_{{k}_{1},\overline{\sigma}} \rangle - \langle {N}_{{k}_{1}-{k}_{2}+k,\overline{\sigma}} \rangle \right) \left[\sum_{{k}^{'}_{1}}{{\Gamma}_{2}({k}^{'}_{1},{k}_{2},k,\omega)}\right] \label{gamma3}
 \raisetag{1pt}
 \end{gather}

 Equation of motion for ${\Gamma}_{4}({k}_{1},{k}_{2},k,\omega)$: 
 
\begin{eqnarray}
\omega  {\Gamma}_{4}({k}_{1},{k}_{2},k,\omega) & \cong & \frac{{\delta}_{{k}_{2},k}}{2\pi}\langle {N}_{{k}_{1},\overline{\sigma}} \rangle +    \left[ {E}_{d}({k}_{1}-{k}_{2}+k) - {E}_{d}({k}_{1}) + {E}_{d}({k}_{2}) + U(n_{d} + 1)\right] {\Gamma}_{4}({k}_{1},{k}_{2},k,\omega) \nonumber  \\
 & + & \{ 2V\langle {N}_{{k}_{1},\overline{\sigma}}\rangle{n}_{c} + (2+\langle {N}_{{k}_{1},\overline{\sigma}} \rangle)Un_{d}+ U(\langle {N}_{{k}_{1},\overline{\sigma}} \rangle - \langle {N}_{{k}_{1}-{k}_{2}+k,\overline{\sigma}} \rangle)\} {F}_{{\sigma}}(k, \omega) \nonumber \\
 & +  & \frac{V}{N} \left(\langle {N}_{{k}_{1},\overline{\sigma}} \rangle - \langle {N}_{{k}_{1}-{k}_{2}+k,\overline{\sigma}} \rangle \right) \left[\sum_{{k}^{'}_{1}}{{\Gamma}_{5}({k}^{'}_{1},{k}_{2},k,\omega)} + \sum_{{k}^{'}_{1}}{{\Gamma}_{6}({k}^{'}_{1},{k}_{2},k,\omega)}\right] \label{gamma4}
 \end{eqnarray}

 Equation of motion for ${\Gamma}_{5}({k}_{1},{k}_{2},k,\omega)$:

 \begin{gather}
\omega  {\Gamma}_{5}({k}_{1},{k}_{2},k,\omega) \cong  \frac{{\delta}_{{k}_{2},k}}{2\pi}\langle {n}_{{k}_{1},{\sigma}} \rangle +  \left[ {E}_{c}({k}_{1}-{k}_{2}+k) - {E}_{c}({k}_{1}) + {E}_{d}({k}_{2}) + Un_{d} + 2{n}_{c}(U+V) - V \right] {\Gamma}_{5}({k}_{1},{k}_{2},k,\omega) \nonumber \\
 + \{ (U-V)\langle {n}_{{k}_{1},{\sigma}}\rangle + 2V\langle {n}_{{k}_{1},{\sigma}} \rangle{n}_{c}- V(\langle {n}_{{k}_{1},{\sigma}} \rangle - \langle {n}_{{k}_{1}-{k}_{2}+k,{\sigma}} \rangle) +V \} {F}_{{\sigma}}(k, \omega) \nonumber \\
 +  \frac{V}{N} \left(\langle {n}_{{k}_{1},{\sigma}} \rangle - \langle {n}_{{k}_{1}-{k}_{2}+k,{\sigma}} \rangle \right) \left[\sum_{{k}^{'}_{1}}{{\Gamma}_{4}({k}^{'}_{1},{k}_{2},k,\omega)} \right] + \frac{U}{N} \left(\langle {n}_{{k}_{1},{\sigma}} \rangle - \langle {n}_{{k}_{1}-{k}_{2}+k,{\sigma}} \rangle \right) \left[\sum_{{k}^{'}_{1}}{{\Gamma}_{6}({k}^{'}_{1},{k}_{2},k,\omega)}\right] \label{gamma5}
 \raisetag{1pt}
 \end{gather}
 
 Equation of motion for ${\Gamma}_{6}({k}_{1},{k}_{2},k,\omega)$: 

\begin{gather}
\omega  {\Gamma}_{6}({k}_{1},{k}_{2},k,\omega) \cong  \frac{{\delta}_{{k}_{2},k}}{2\pi}\langle {n}_{{k}_{1},\overline{\sigma}} \rangle +
              \left[ {E}_{c}({k}_{1}-{k}_{2}+k) - {E}_{c}({k}_{1}) + {E}_{d}({k}_{2}) + Un_{d} + 2V{n}_{c} + V \right] {\Gamma}_{6}({k}_{1},{k}_{2},k,\omega)  \nonumber \\
 + \{ U\langle {n}_{{k}_{1},\overline{\sigma}}\rangle n_{d} + V(1 - \langle {N}_{{k}_{1},\overline{\sigma}} \rangle -  \langle {N}_{{k}_{1}-{k}_{2}+k,\overline{\sigma}} \rangle){n}_{c} + V(\langle {N}_{{k}_{1},\overline{\sigma}} \rangle \} {F}_{{\sigma}}(k, \omega) \nonumber \\
 +  \frac{V}{N} \left(\langle {n}_{{k}_{1},\overline{\sigma}} \rangle - \langle {n}_{{k}_{1}-{k}_{2}+k,\overline{\sigma}} \rangle \right) \left[\sum_{{k}^{'}_{1}}{{\Gamma}_{4}({k}^{'}_{1},{k}_{2},k,\omega)} \right] +  \frac{U}{N} \left(\langle {n}_{{k}_{1},\overline{\sigma}} \rangle - \langle {n}_{{k}_{1}-{k}_{2}+k,\overline{\sigma}} \rangle \right) \left[\sum_{{k}^{'}_{1}}{{\Gamma}_{5}({k}^{'}_{1},{k}_{2},k,\omega)}\right] \label{gamma6}
 \raisetag{1pt}
 \end{gather}

Notice that this RPA-like approximation involves a second-order closed set of equations  
 for $G_{\sigma}(k,\omega)$, and double k-index summations of  $\Gamma_{1}$, $\Gamma_{2}$ and $\Gamma_{3}$;                                   
  and an analogous one for 
 $F_{\sigma}(k,\omega)$, $\Gamma_{4}$, $\Gamma_{5}$ and $\Gamma_{6}$. Below we explain how we could solve these two sets of coupled equations 
 of motion, and detail the Green's functions $G(\vec{k},\omega)$ and $F(\vec{k},\omega)$ obtained, required 
to calculate the spectral density and total density of states results presented in Section \ref{results}.
 
\subsubsection{Solution of the equations of motion determining ${G}_{{\sigma}}(k, \omega)$.}

The keypoint to close the  EOM system at this second-order level, enabling to determine  $G_{\sigma}(k,\omega)$ 
from Eq. \ref{eqG} using the RPA-coupled set of   Eqs. \ref{gamma1},\ref{gamma2} an d \ref{gamma3}, 
is that only  BZ summations over $k_1$ and $ k_2$  of the ${\Gamma}_{i}({k}_{1},{k}_{2},k)$ (i=1,3) functions 
appear in    Eq. \ref{eqG}. So that, appropriately (double) summing over the BZ  the set of coupled set of equations, 
one can effectively extract $G_{\sigma}(k,\omega)$ at  a fixed $k$ as we explain in the following.

  It is useful to introduce the following notation: 
\begin{equation}
 \Phi_{i}(k_{2},k,\omega) \equiv \sum_{k_{1}}{\Gamma_{i}(k_{1},k_{2},k,\omega)}    \quad , \quad i= 1,3  
 \end{equation}

\noindent which allows to rewrite Eqs.\ref{gamma1}, \ref{gamma2} and \ref{gamma3}  as follows:
\begin{eqnarray}\label{sub1}
{\Gamma}_{1}({k}_{1},{k}_{2},k,\omega) &  =   &A_{1} + B_{1}{G}_{{\sigma}}(k, \omega) + C_{1}\left[   \Phi_{2}(k_{2},k,\omega) + \Phi_{3}(k_{2},k,\omega) \right] \nonumber \\
{\Gamma}_{2}({k}_{1},{k}_{2},k,\omega) &  =   &A_{2} + B_{2}{G}_{{\sigma}}(k, \omega) + C_{2}\left[  \left( \frac{V}{N} \right) \Phi_{1}(k_{2},k,\omega) +  \left( \frac{U}{N} \right)\Phi_{3}(k_{2},k,\omega) \right] \nonumber\\
{\Gamma}_{3}({k}_{1},{k}_{2},k,\omega) &  =   &A_{3} + B_{3}{G}_{{\sigma}}(k, \omega) + C_{3}\left[  \left( \frac{V}{N} \right) \Phi_{1}(k_{2},k,\omega) +  \left( \frac{U}{N} \right)\Phi_{2}(k_{2},k,\omega) \right] 
\end{eqnarray}

\noindent where the coefficients $A_{i}$, $B_{i}$ and $C_{i}$ are:

\begin{eqnarray*}
A_{1} &  =  & \frac{\langle {n}_{{k}_{1},\overline{\sigma}} \rangle}{2\pi \left[ \omega - E_{c}(k_{2})-U(n_{c}+1) \right]} \\
A_{2} &  =  & \frac{\langle {N}_{{k}_{1},{\sigma}} \rangle}{2\pi \left[ \omega - E_{c}(k_{2})-Un_{c}-2n_{d}(U+V) + V \right]} \\
A_{3} &  =  & \frac{\langle {N}_{{k}_{1},\overline{\sigma}} \rangle}{2\pi \left[ \omega - E_{c}(k_{2})-Un_{c}-2Vn_{d} - V \right]}
\end{eqnarray*}

\begin{eqnarray*}
B_{1} &  =  &  \frac{\{ 2V\langle {n}_{{k}_{1},\overline{\sigma}}\rangle n_{d} + (2+\langle {n}_{{k}_{1},\overline{\sigma}} \rangle)U{n}_{c}+ U(\langle {n}_{{k}_{1},\overline{\sigma}} \rangle - \langle {n}_{{k}_{1}-{k}_{2}+k,\overline{\sigma}} \rangle)\}}{\omega - \omega_{1}} \\
B_{2} &  =  & \frac{\{ (U-V)\langle {N}_{{k}_{1},{\sigma}}\rangle + 2V\langle {N}_{{k}_{1},{\sigma}} \rangle n_{d}- V(\langle {N}_{{k}_{1},{\sigma}} \rangle - \langle N_{{k}_{1}-{k}_{2}+k,{\sigma}} \rangle) +V \}}{\omega - \omega_{2}} \\
B_{3} &  =  & \frac{\{ U\langle {N}_{{k}_{1},\overline{\sigma}}\rangle {n}_{c} + V(1 - \langle {n}_{{k}_{1},\overline{\sigma}} \rangle -  \langle {n}_{{k}_{1}-{k}_{2}+k,\overline{\sigma}} \rangle)n_{d} + V(\langle {n}_{{k}_{1},\overline{\sigma}} \rangle \}}{\omega - \omega_{3}}
\end{eqnarray*}

\begin{eqnarray*}
C_{1} &  =  &  \frac{\frac{V}{N} \left[\langle {n}_{{k}_{1},\overline{\sigma}} \rangle - \langle {n}_{{k}_{1}-{k}_{2}+k,\overline{\sigma}} \rangle \right]}{\omega - \omega_{1}} \\
C_{2} &  =  & \frac{\left[\langle {N}_{{k}_{1},{\sigma}} \rangle - \langle {N}_{{k}_{1}-{k}_{2}+k,{\sigma}} \rangle \right]}{\omega - \omega_{2}} \\
C_{3} &  =  & \frac{\left[\langle {N}_{{k}_{1},\overline{\sigma}} \rangle - \langle {N}_{{k}_{1}-{k}_{2}+k,\overline{\sigma}} \rangle \right] }{\omega - \omega_{3}}
\end{eqnarray*}

\noindent and, 

\begin{eqnarray*}
\omega_{1} &  =  &  \left[ {E}_{c}({k}_{1}-{k}_{2}+k) - {E}_{c}({k}_{1}) + {E}_{c}({k}_{2}) + U({n}_{c} + 1) \right] \\
\omega_{2} &  =  & \left[ {E}_{d}({k}_{1}-{k}_{2}+k) - {E}_{d}({k}_{1}) + {E}_{c}({k}_{2}) + U{n}_{c} + 2n_{d}(U+V) - V \right] \\
\omega_{3} &  =  &  \left[ {E}_{d}({k}_{1}-{k}_{2}+k) - {E}_{d}({k}_{1}) + {E}_{c}({k}_{2}) + U{n}_{c} + 2Vn_{d} + V \right]
\end{eqnarray*}


Notice that if  Eqs.\ref{sub1} are BZ summed  over $k_1$, each of the  $ \Phi_{i}(k_{2},k) $ can be derived in terms of   $G_{\sigma}(k,\omega)$
and coefficients involving the temperature dependant-fillings, the effective band structure and the model's correlation parameters.        

Meanwhile,
if one introduces the notation: 
\begin{equation} \label{defin-tita}
\Theta_{i}(k) \equiv  \sum_{k_{1}, k_{2}} {\Gamma_{i}(k_{1},k_{2},k,\omega)}  \equiv  \sum_{k_{2}}  \Phi_{i}(k_{2},k,\omega)  \quad , \quad i= 1,3  
 \end{equation}

 \noindent Eq. \ref{eqG} can be rewritten as:
 
 \begin{align}
   \left[\omega - E_{c}(k) \right] G_{\sigma}(k,\omega)  =  \frac{1}{2\pi} +  
   {\left[\frac{U}{N} \Theta_{1}(k,\omega) + \frac{V}{N} \Theta_{2}(k,\omega)  + \frac{V}{N} \Theta_{3}(k,\omega) \right] }   
 \label{eqG-tita}
\raisetag{1pt}
  \end{align}
 
Performing the $k_2 $ summations in Eq. \ref{defin-tita}, all $\Theta_{i}(k)$ can be expressed in terms of   $G_{\sigma}(k,\omega)$
and coefficients involving the temperature dependant-fillings, the effective   and the model's correlation parameters, 
so that one can finally solve for $G_{\sigma}(k,\omega)$.   We find the following  expression for G($\vec{k}$, $\omega$), 
related to the $c$-electrons in our model, in this RPA- level approximation: 


\begin{equation}\label{G}
G^{RPA}_{{\sigma}}(k, \omega) \cong \frac{ \frac{1}{2\pi} + \sum_{k_{2}}  {\left \{ \frac{U}{N}\left(\frac{Y_{1}}{X_{1}}\right)  + \frac{V}{N}\left[ \frac{Y_{2}}{X_{2}} + [A_{3}]_{k_{1}} +  [C_{3}]_{k_{1}}\left(\frac{Y_{1}}{X_{1}}+\frac{Y_{2}}{X_{2}}\right)     \right]  \right\} }     }{ \omega - E_{c}(k) -\sum_{k_{2}}{\left \{\frac{U}{N}\left(\frac{Z_{1}}{X_{1}}\right)+\frac{V}{N} \left[ \frac{Z_{2}}{X_{2}}+[B_{3}]_{k_{1}} +[C_{3}]_{k_{1}} \left( \frac{Z_{1}}{X_{1}}+\frac{Z_{2}}{X_{2}} \right) \right] \right\}} }
\end{equation}

\noindent where, denoting: $[A]_{k_{1}} \equiv \sum_{k_{1}}{A}$, one has: 

\begin{eqnarray*}
X_{1}(k_{2},k) & = &\left(1-\frac{U}{N}[C_{2}]_{k_{1}}[C_{3}]_{k_{1}}\right) \left( 1-[C_{1}]_{k_{1}}[C_{3}]_{k_{1}}  \right) - [C_{1}]_{k_{1}}\left( \frac{U}{N}[C_{3}]_{k_{1}} + 1 \right) \left( \frac{U}{N}[C_{3}]_{k_{1}} + \frac{V}{N} \right) [C_{2}]_{k_{1}}\\
Y_{1}(k_{2},k) & = &\left(1-\frac{U}{N}[C_{2}]_{k_{1}}[C_{3}]_{k_{1}}\right) \left( [A_{1}]_{k_{1}} +  [A_{3}]_{k_{1}}[C_{1}]_{k_{1}} \right) + \left( [A_{2}]_{k_{1}} + \frac{U}{N}[C_{2}]_{k_{1}}[A_{3}]_{k_{1}} \right) [C_{1}]_{k_{1}} \left( \frac{U}{N}[C_{3}]_{k_{1}} + 1 \right)\\
Z_{1}(k_{2},k) & = &\left(1-\frac{U}{N}[C_{2}]_{k_{1}}[C_{3}]_{k_{1}}\right)\left( [B_{1}]_{k_{1}} +  [B_{3}]_{k_{1}}[C_{1}]_{k_{1}} \right) + [B_{2}]_{k_{1}} + \frac{U}{N}[C_{2}]_{k_{1}}[B_{3}]_{k_{1}}
\end{eqnarray*}

\begin{eqnarray*}
 X_{2}(k_{2},k)& = &\left(1-[C_{1}]_{k_{1}}[C_{3}]_{k_{1}}\right) \left( 1-\frac{U}{N}[C_{2}]_{k_{1}}[C_{3}]_{k_{1}}  \right) - \left(\frac{U}{N}[C_{3}]_{k_{1}} + V\right)[C_{2}]_{k_{1}}  [C_{1}]_{k_{1}} \left([C_{3}]_{k_{1}} + 1\right)\\
 Y_{2}(k_{2},k) &= &\left(1-[C_{1}]_{k_{1}}[C_{3}]_{k_{1}}\right) \left( [A_{2}]_{k_{1}} + \frac{U}{N} [C_{2}]_{k_{1}}[A_{3}]_{k_{1}} \right) + \left( \frac{U}{N} [C_{3}]_{k_{1}} + V \right) [C_{2}]_{k_{1}}\left( [A_{1}]_{k_{1}} +[A_{3}]_{k_{1}}[C_{1}]_{k_{1}}\right)\\
 Z_{2}(k_{2},k)& =& \left(1-[C_{1}]_{k_{1}}[C_{3}]_{k_{1}}\right) \left( [B_{2}]_{k_{1}} + \frac{U}{N} [C_{2}]_{k_{1}}[B_{3}]_{k_{1}} \right) + \left( \frac{U}{N} [C_{3}]_{k_{1}} + V \right) [C_{2}]_{k_{1}}\left( [B_{1}]_{k_{1}} +[B_{3}]_{k_{1}}[C_{1}]_{k_{1}}\right)
\end{eqnarray*}

\subsubsection{Solution of the equations of motion determining ${F}_{{\sigma}}(k, \omega)$.}

We proceeded likewise for the subset of equations for  $\Gamma_{i}$ ($i= 4...6)$  coupled to $F_{\sigma}(k,\omega)$:

\begin{eqnarray}\label{sub2}
{\Gamma}_{4}({k}_{1},{k}_{2},k,\omega) &  =   &A^{*}_{1} + B^{*}_{1} {F}_{{\sigma}}(k, \omega) + C^{*}_{1}\left[   \Phi_{5}(k_{2},k,\omega) + \Phi_{6}(k_{2},k,\omega) \right] \nonumber\\
{\Gamma}_{5}({k}_{1},{k}_{2},k,\omega) &  =   &A^{*}_{2} + B^{*}_{2} {F}_{{\sigma}}(k, \omega) + C^{*}_{2}\left[  \left( \frac{V}{N} \right) \Phi_{4}(k_{2},k,\omega) +  \left( \frac{U}{N} \right)\Phi_{6}(k_{2},k,\omega) \right] \nonumber\\
{\Gamma}_{6}({k}_{1},{k}_{2},k,\omega) &  =   &A^{*}_{3} + B^{*}_{3} {F}_{{\sigma}}(k, \omega) + C^{*}_{3}\left[  \left( \frac{V}{N} \right) \Phi_{4}(k_{2},k,\omega) +  \left( \frac{U}{N} \right)\Phi_{5}(k_{2},k,\omega) \right] 
\end{eqnarray}

\noindent where,

\begin{eqnarray*}
A^{*}_{1} &  =  & \frac{\langle {N}_{{k}_{1},\overline{\sigma}} \rangle}{2\pi \left[ \omega - E_{d}(k_{2})-U(n_{d}+1) \right]} \\
A^{*}_{2} &  =  & \frac{\langle {n}_{{k}_{1},{\sigma}} \rangle}{2\pi \left[ \omega - E_{d}(k_{2})-Un_{d}-2n_{c}(U+V) + V \right]} \\
A^{*}_{3} &  =  & \frac{\langle {n}_{{k}_{1},\overline{\sigma}} \rangle}{2\pi \left[ \omega - E_{d}(k_{2})-Un_{d}-2Vn_{c} - V \right]}
\end{eqnarray*}

\begin{eqnarray*}
B^{*}_{1} &  =  &  \frac{\{ 2V\langle {N}_{{k}_{1},\overline{\sigma}}\rangle{n}_{c} + (2+\langle {N}_{{k}_{1},\overline{\sigma}} \rangle)Un_{d}+ U(\langle {N}_{{k}_{1},\overline{\sigma}} \rangle - \langle {N}_{{k}_{1}-{k}_{2}+k,\overline{\sigma}} \rangle)\}}{\omega - \omega^{*}_{1}} \\
B^{*}_{2} &  =  & \frac{\{ (U-V)\langle {N}_{{k}_{1},{\sigma}}\rangle + 2V\langle {n}_{{k}_{1},{\sigma}} \rangle {n}_{c}- V(\langle {n}_{{k}_{1},{\sigma}} \rangle - \langle n_{{k}_{1}-{k}_{2}+k,{\sigma}} \rangle) +V \}}{\omega - \omega^{*}_{2}} \\
B^{*}_{3} &  =  & \frac{\{ U\langle {n}_{{k}_{1},\overline{\sigma}}\rangle n_{d} + V(1 - \langle {N}_{{k}_{1},\overline{\sigma}} \rangle -  \langle {N}_{{k}_{1}-{k}_{2}+k,\overline{\sigma}} \rangle){n}_{c} + V(\langle {N}_{{k}_{1},\overline{\sigma}} \rangle \}}{\omega - \omega^{*}_{3}}
\end{eqnarray*}

\begin{eqnarray*}
C^{*}_{1} &  =  & \frac{V}{N} \frac{ \left[\langle {N}_{{k}_{1},\overline{\sigma}} \rangle - \langle {N}_{{k}_{1}-{k}_{2}+k,\overline{\sigma}} \rangle \right]}{\omega - \omega^{*}_{1}} \\
C^{*}_{2} &  =  & \frac{\left[\langle {n}_{{k}_{1},{\sigma}} \rangle - \langle {n}_{{k}_{1}-{k}_{2}+k,{\sigma}} \rangle \right]}{\omega - \omega^{*}_{2}} \\
C^{*}_{3} &  =  & \frac{\left[\langle {n}_{{k}_{1},\overline{\sigma}} \rangle - \langle {n}_{{k}_{1}-{k}_{2}+k,\overline{\sigma}} \rangle \right] }{\omega - \omega^{*}_{3}}
\end{eqnarray*}

\begin{eqnarray*}
\omega^{*}_{1} &  =  &  \left[ {E}_{d}({k}_{1}-{k}_{2}+k) - {E}_{d}({k}_{1}) + {E}_{d}({k}_{2}) + U(n_{d} + 1) \right] \\
\omega^{*}_{2} &  =  & \left[ {E}_{c}({k}_{1}-{k}_{2}+k) - {E}_{c}({k}_{1}) + {E}_{d}({k}_{2}) + Un_{d} + 2{n}_{c}(U+V) - V \right] \\
\omega^{*}_{3} &  =  &  \left[ {E}_{c}({k}_{1}-{k}_{2}+k) - {E}_{c}({k}_{1}) + {E}_{d}({k}_{2}) + Un_{d} + 2V{n}_{c} + V \right]
\end{eqnarray*}


Carrying out the summations over $\vec{k_{1}}$ and $\vec{k_{2}}$ BZ points in the set of equations in Eq. \ref{sub2} and Eq.\ref{eqF}, 
we analogously find an expression for the Green's function $F_{\sigma}(k,\omega)$ corresponding
to the d-electrons, in this RPA-level approximation:


\begin{equation}\label{F}
F^{RPA}_{{\sigma}}(k, \omega) \cong \frac{ \frac{1}{2\pi} + \sum_{k_{2}}  {\left \{ \frac{U}{N}\left(\frac{Y^{*}_{1}}{X^{*}_{1}}\right)  + \frac{V}{N}\left[ \frac{Y^{*}_{2}}{X^{*}_{2}} + [A^{*}_{3}]_{k_{1}}  +  [C^{*}_{3}]_{k_{1}}\left(\frac{Y^{*}_{1}}{X^{*}_{1}}+ \frac{Y^{*}_{2}}{X^{*}_{2}}\right)     \right]  \right\} }     }{ \omega - E_{d}(k) -\sum_{k_{2}}{\left \{\frac{U}{N}\left(\frac{Z^{*}_{1}}{X^{*}_{1}}\right)+\frac{V}{N} \left[ \frac{Z^{*}_{2}}{X^{*}_{2}}+[B^{*}_{3}]_{k_{1}} +[C^{*}_{3}]_{k_{1}} \left( \frac{Z^{*}_{1}}{X^{*}_{1}}+\frac{Z^{*}_{2}}{X^{*}_{2}} \right) \right] \right\}} }
\end{equation}

\noindent where:

\begin{eqnarray*}
X^{*}_{1}(k_{2},k) & = &\left(1-\frac{U}{N}[C^{*}_{2}]_{k_{1}}[C^{*}_{3}]_{k_{1}}\right) \left( 1-[C^{*}_{1}]_{k_{1}}[C^{*}_{3}]_{k_{1}}  \right) - [C^{*}_{1}]_{k_{1}}\left( \frac{U}{N}[C^{*}_{3}]_{k_{1}} + 1 \right) \left( \frac{U}{N}[C^{*}_{3}]_{k_{1}} + \frac{V}{N} \right) [C^{                                                                        *}_{2}]_{k_{1}}\\
Y^{*}_{1}(k_{2},k) & = &\left(1-\frac{U}{N}[C^{*}_{2}]_{k_{1}}[C^{*}_{3}]_{k_{1}}\right) \left( [A^{*}_{1}]_{k_{1}} +  [A^{*}_{3}]_{k_{1}}[C^{*}_{1}]_{k_{1}} \right) + \left( [A^{*}_{2}]_{k_{1}} + \frac{U}{N}[C^{*}_{2}]_{k_{1}}[A^{*}_{3}]_{k_{1}} \right) [C^{*}_{1}]_{k_{1}}\left( \frac{U}{N}[C^{*}_{3}]_{k_{1}} + 1 \right)\\
Z^{*}_{1}(k_{2},k) & = &\left(1-\frac{U}{N}[C^{*}_{2}]_{k_{1}}[C^{*}_{3}]_{k_{1}}\right)\left( [B^{*}_{1}]_{k_{1}} +  [B^{*}_{3}]_{k_{1}}[C^{*}_{1}]_{k_{1}} \right) + [B^{*}_{2}]_{k_{1}} + \frac{U}{N}[C^{*}_{2}]_{k_{1}}[B^{*}_{3}]_{k_{1}}
\end{eqnarray*}

\begin{eqnarray*}
 X^{*}_{2}(k_{2},k)& = &\left(1-[C^{*}_{1}]_{k_{1}}[C^{*}_{3}]_{k_{1}}\right) \left( 1-\frac{U}{N}[C^{*}_{2}]_{k_{1}}[C^{*}_{3}]_{k_{1}}  \right) - \left(\frac{U}{N}[C^{*}_{3}]_{k_{1}} + V\right)[C^{*}_{2}]_{k_{1}}  [C^{*}_{1}]_{k_{1}} \left([C^{*}_{3}]_{k_{1}} + 1\right)\\
 Y^{*}_{2}(k_{2},k) &= &\left(1-[C^{*}_{1}]_{k_{1}}[C^{*}_{3}]_{k_{1}}\right) \left( [A^{*}_{2}]_{k_{1}} + \frac{U}{N} [C^{*}_{2}]_{k_{1}}[A^{*}_{3}]_{k_{1}} \right) + \left( \frac{U}{N} [C^{*}_{3}]_{k_{1}} + V \right) [C^{*}_{2}]_{k_{1}}\left( [A^{*}_{1}]_{k_{1}} +[A^{*}_{3}]_{k_{1}}[C^{*}_{1}]_{k_{1}}\right)\\
 Z^{*}_{2}(k_{2},k)& =& \left(1-[C^{*}_{1}]_{k_{1}}[C^{*}_{3}]_{k_{1}}\right) \left( [B^{*}_{2}]_{k_{1}} + \frac{U}{N} [C^{*}_{2}]_{k_{1}}[B_{3}]_{k_{1}} \right) + \left( \frac{U}{N} [C^{*}_{3}]_{k_{1}} + V \right) [C^{*}_{2}]_{k_{1}}\left( [B^{*}_{1}]_{k_{1}} +[B^{*}_{3}]_{k_{1}}[C^{*}_{1}]_{k_{1}}\right)
\end{eqnarray*}

\end{widetext}

 \section{Analysis of the k-dependent self-energy}
 \label{apen2}

To better assess the nature of the second order (RPA-like) approximation which we 
used to decouple the equations of motion and determine the relevant Green's functions, 
yielding the spectral density and total density of states of our model, 
in the following  we present an analysis of the self-energies for $c$ and $d$ electrons
obtained in our approach, focusing on the k-dependence and the temperature dependence.
 
Using Dyson's equation, one can determine the second order self-energy related to each electron band 
as follows:
\begin{equation}
 \Sigma_{c}(k,\omega) = G^{0}(k,\omega)^{-1} -  G^{RPA}(k,\omega)^{-1}
\end{equation}

\begin{equation}
 \Sigma_{d}(k,\omega) = F^{0}(k,\omega)^{-1} -  F^{RPA}(k,\omega)^{-1}
\end{equation}

\noindent where, $G^{0}(k,\omega)$ and $F^{0}(k,\omega)$ represent the retarded non-interacting 
Green's functions corresponding to the two bare effective bands:  for the 
uncorrelated c- and d-bands, respectively,   $ G^{0}(k,\omega)^{-1} = \omega - E_c ( \vec{k} ) $ and $ F^{0}(k,\omega)^{-1} = \omega - E_d ( \vec{k} ) $. Meanwhile:  $G^{RPA}(k,\omega)$ and $F^{RPA}(k,\omega)$  were defined  in Eqs. A22 and A24 of Appendix A.

First, notice that the self-energy in Hartree-Fock approximation for the c- and d-orbital  Green's functions (Eqs. A4 and A6)
is independent of crystal momentum k and energy. 

Therefore,  in this appendix  we present  our results obtained evaluating in  second order  level of approximation (as detailed in Appendix \ref{apen1}) 
the real and imaginary parts of the self-energies defined above, corresponding to the retarded electron Green's functions of eqs. A22 and A24. 
Our results confirm that in our RPA-like approach,  the self-energies exhibit clear renormalization effects due to the correlations included:   
with non-trivial crystal momentum, temperature, doping, and energy dependence.

  \begin{figure}[h!]
  \begin{center}
  \includegraphics[width=8.6cm, height=8.6cm]{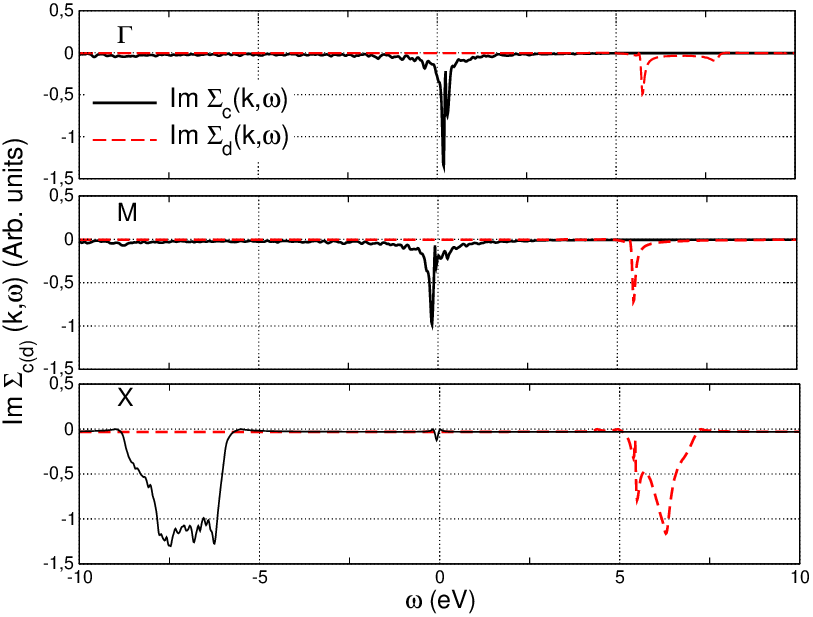}
     \caption[]{Imaginary part of the  c- and d-band self-energies as a function of energy (w.r. to the Fermi level),  at the following symmetry points of the 
     square lattice Brillouin zone: $\Gamma$ ( $\vec{k}=(0,0)$), $M$( $\vec{k}=(\pi,\pi)$) and $X$ ( $\vec{k}=(\pi,0)$). 
     Parameters:  $U = V = 3.5$, $n_{c}=1.45$ and $n_{d}=0.56$, $T = 20K$. Bare effective band tight-binding parameters as in Ref. \onlinecite{raghu}  }
  \label{fig1}
  \end{center}
  \end{figure}

In the following, our results will be presented in a series of figures, in particular 
  exhibiting  the diversity of the temperature dependence which is  obtained at different k-points in the Brillouin zone.
As general consistency checks of our results, notice:  the correct (negative) sign exhibited by the imaginary parts of the electron self-energies, 
while the real parts of the self-energies, which represent the energy shifts renormalizing the bare electron energies in our approximation,
 are correctly Kramers-Kronig related to the respective imaginary parts.

\subsection{Momentum dependence of $\Sigma_{c}(k,\omega)$ and $\Sigma_{d}(k,\omega)$  }

First, we exemplify the momentum dependence of the imaginary part of
the two self-energies $\Sigma_{c}(k,\omega)$ and $\Sigma_{d}(k,\omega)$ in our approximation.
In Fig. \ref{fig1}, we show the imaginary parts of the two self-energies at the three relevant high symmetry BZ points of the square lattice, 
usually probed by ARPES. The upper  two panels  show that at  the Brillouin zone centre ($\Gamma$) and at  $M$, 
 the c-band is mainly renormalized  around the Fermi level,  while the d-band is mostly renormalized  far away  from the Fermi level, 
 at energies close to the bare band edges  ($\omega>5$eV). The lower panel depicts the imaginary parts of the two self-energies 
   at $X$ ($ \vec{k}=(\pi,0)$): which noticeably differs from the previous cases. At $X$ the renormalization  is relevant 
   mostly near the band edges of the bare electronic structure, while near the Fermi level the renormalization at $X$ is not significant.
  
In the rest of this appendix, we will focus on the renormalization of the energies obtained near the Fermi level, 
and  therefore center our  discussion  on $\Sigma_{c}(\vec{k},\omega)$.

 \begin{figure}[h!]
  \begin{center}
  \includegraphics[width=8.6cm]{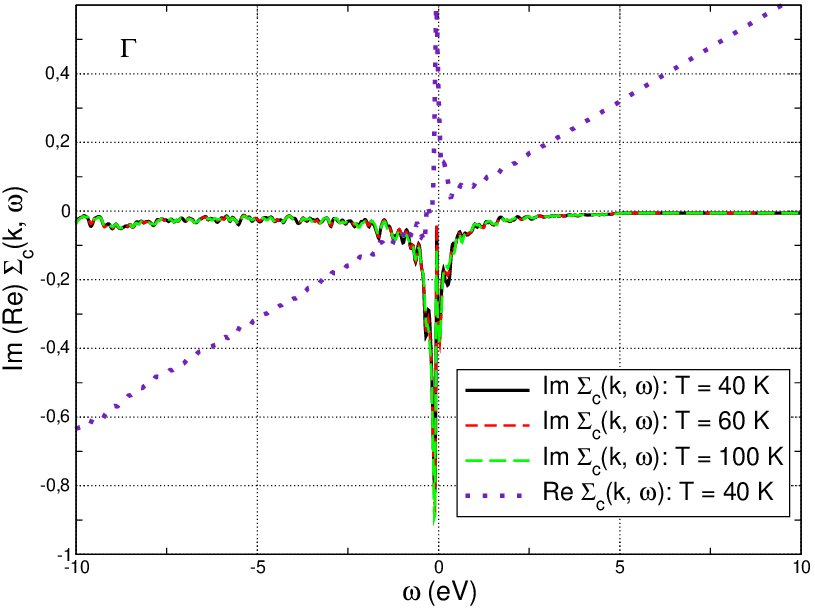}
     \caption[]{Temperature dependence of the imaginary part of $\Sigma_{c}(k,\omega)$
     at $\Gamma$. $T$ as labelled in the figure, other parameters as in Fig. \ref{fig1}.}
  \label{fig4}
  \end{center}
  \end{figure}

  \begin{figure}[t]
  \begin{center}
  \includegraphics[width=8.6cm, height=6cm]{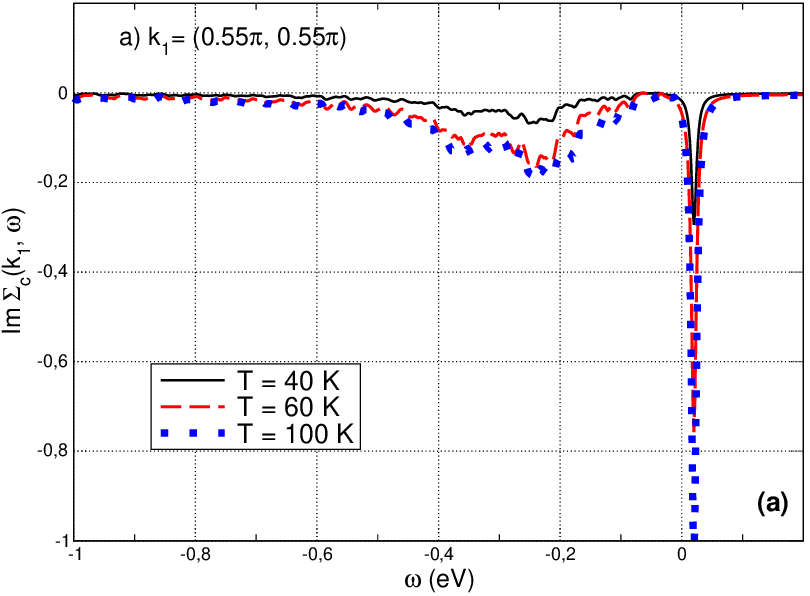}
   \includegraphics[width=8.6cm, height=6cm]{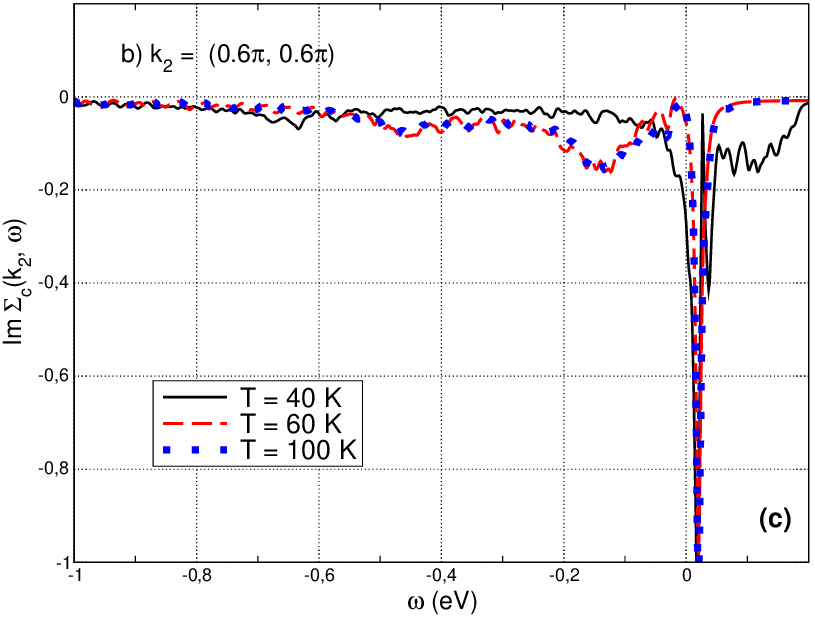}
  \includegraphics[width=8.6cm, height=6cm]{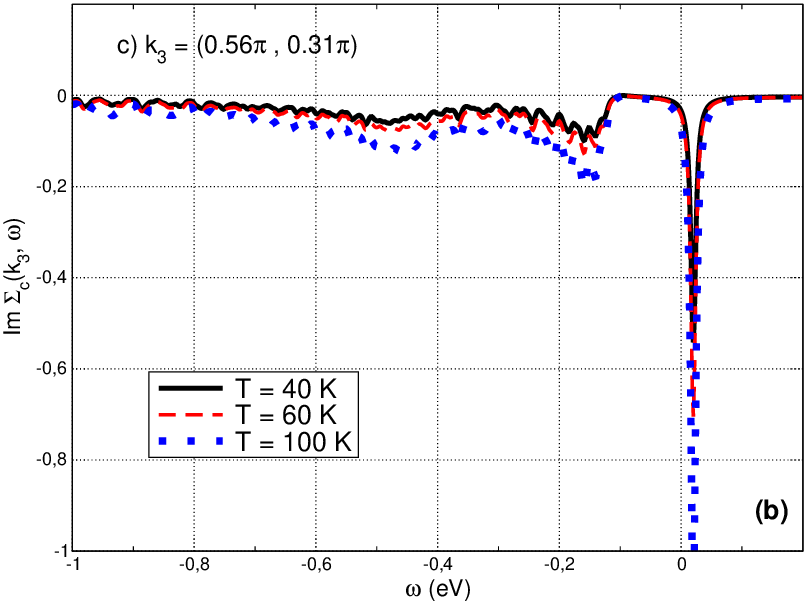}
     \caption[]{Temperature dependence of the imaginary part of $\Sigma_{c}(k,\omega)$
     at the BZ points respectively indicated in the plots.
      $T$ as labelled in the figure, other parameters as in Fig. \ref{fig1}.}
  \label{fig8}
  \end{center}
  \end{figure}

 \begin{figure}[t]
  \begin{center}
   \includegraphics[width=8.6cm, height=6cm]{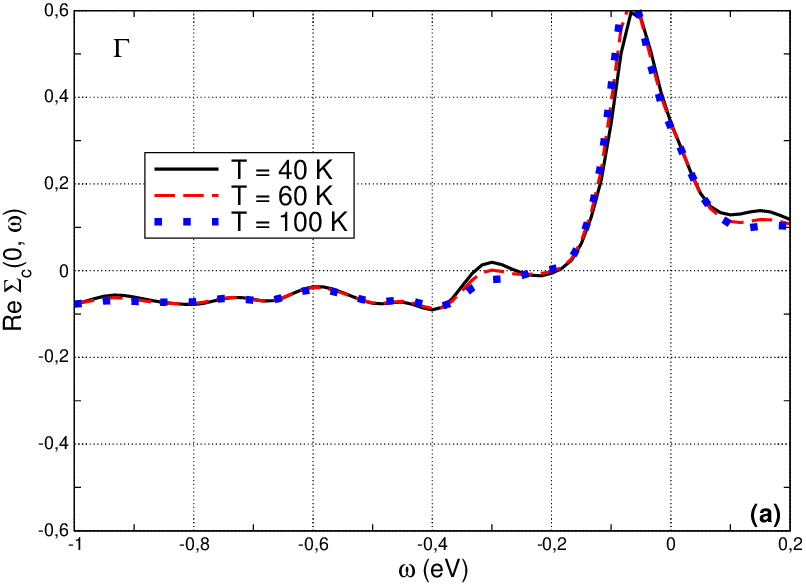}
  \includegraphics[width=8.6cm, height=6cm]{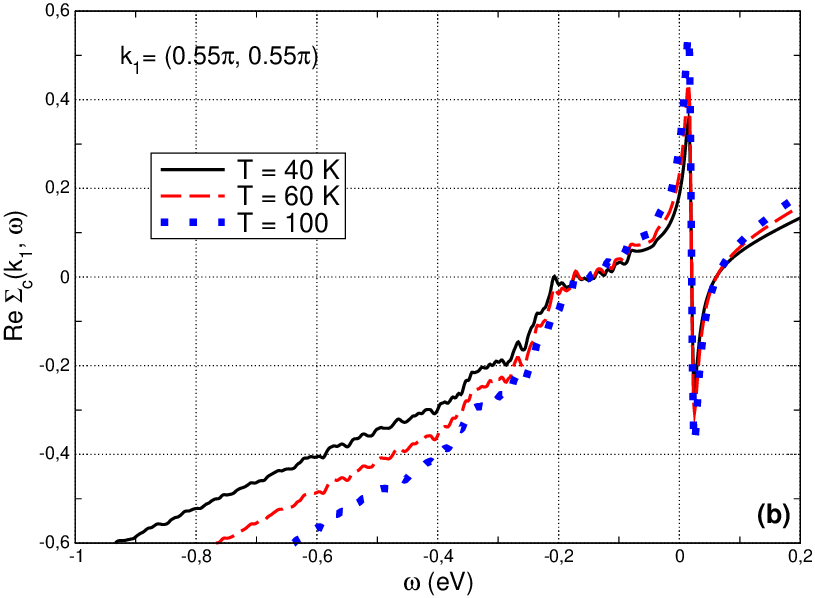}
  \includegraphics[width=8.6cm, height=6cm]{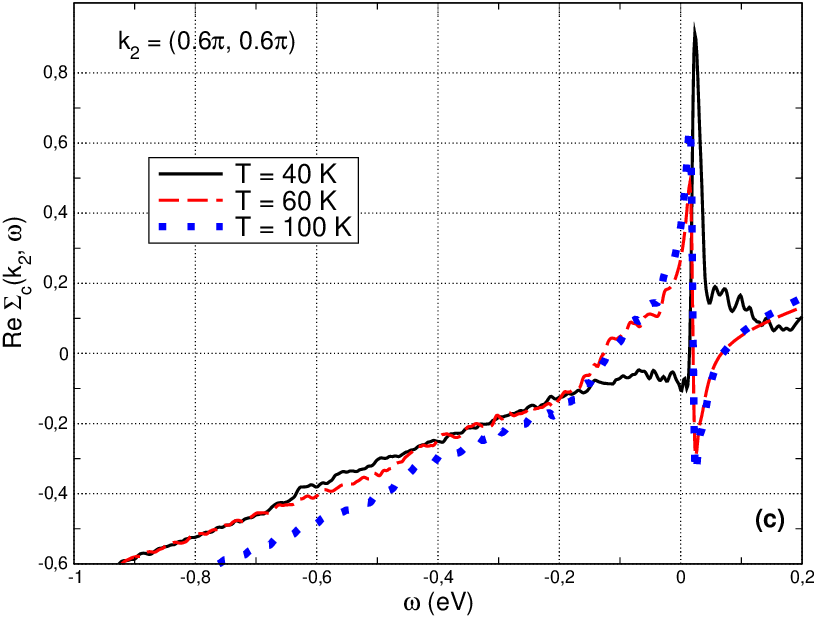}
     \caption[]{Temperature dependence of the real part of  $\Sigma_{c}(k,\omega)$
     at the BZ points respectively indicated in the plots.  $T$ as labelled in the figure, other parameters as in Fig.  \ref{fig1}.}
  \label{real-gamma}
  \end{center}
  \end{figure}

\subsection{Temperature dependence of the self-energy}

 \subsubsection{Imaginary part of the self-energy}
 
 Now, we will focus on the evolution with temperature of the imaginary part of $\Sigma_{c}(\vec{k},\omega)$, in particular 
 fixing $\vec{k}$ at  the relevant high-symmetry points of the BZ studied by  ARPES experiments, and also at other
 BZ points: in which we found that the temperature dependence of the renormalization was larger and non-trivial (not the expected 
 behaviour resulting from $ f_{FD} (\omega) $).
 
 Fig. \ref{fig4} shows that at $\Gamma$ the renormalization  of the self-energy is almost independent of temperature.  We found that also 
 at  $M$ and $X$  the effect of temperature on the renormalization  is almost irrelevant. This negligible temperature dependence of the
   self-energy  at the  high symmetry BZ points explored by ARPES,  $\Gamma$, $M$ and $X$,
    explains the spectral density function results presented in Figs. 10 and 11 of Section \ref{lastsubsection}.

 But from the total density of states results in Fig. 9 of Section \ref{beforelast}, we know that non-trivial relevant temperature-dependent effects indeed appear 
 when the contributions of the whole BZ are taken into account. 
 Therefore, we used a grid to explore the temperature dependence of the  renormalization at different points of the BZ (528 points), 
 and  interestingly  we could identify a number of specific BZ points, not yet probed by ARPES,  where  sizeable  non-trivial  temperature dependent
  renormalization effects are obtained.

Concretely, in Figure \ref{fig8}(a) we exhibit the temperature dependence of the imaginary part of $\Sigma_{c}(k,\omega)$  at $\vec{k} = (0.55\pi,0.55\pi)$: 
a BZ point where we find that,  increasing temperature,   interaction-related renormalization effects
lead to  a redistribution of spectral weight  near the Fermi level,  with temperature dependent peaks evolving 
 in the energy range $[-0.4,-0.2]$ eV. 
 
In  Figure \ref{fig8}(b), we exhibit the temperature dependence of the imaginary part of $\Sigma_{c}(k,\omega)$ at  $\vec{k} = (0.6\pi,0.6\pi)$, 
where the spectral weight redistribution is even larger, involving e.g. a number of states present above the Fermi level at T= 40K ,
 which are pushed below the Fermi level  at higher temperatures, leading to a number of peaks with temperature-dependent location and height
in the energy range $[-0.7,0]$ eV.

Finally, in Figure \ref{fig8}(c)  we exhibit the temperature dependence of the imaginary part of $\Sigma_{c}(k,\omega)$ at $\vec{k} = (0.56\pi,0.31\pi)$. 
Notice that, increasing temperature,  the interaction effects lead to  temperature dependent peaks  now in the energy range $[-0.6,-0.1]$ eV. 
  It would certainly be interesting to investigate these predictions by ARPES.

 \subsubsection{Real part of the self-energy}

  To complete the presentation of the renormalization effects described in our approach, in Figure \ref{real-gamma} we show the evolution with temperature of the real part of $\Sigma_{c}(k,\omega)$  at the BZ centre, where it is negligible,  and at two of the k-points we identified as having  relevant non-trivial temperature dependence, discussed above.  The  results  in Fig. \ref{real-gamma} are the Kramers-Kronig related counterparts 
  of the imaginary part of the self-energies shown in Figs. \ref{fig4}, \ref{fig8}(a) and \ref{fig8}(b) respectively.
  We checked that the renormalizations at $M$ and $X$, as expected, are almost independent of temperature.

\section{Description of doping effects on the electronic structure}
   
  
Here, we complement our discussion of the effects of doping on the electronic structure of Section 3.3, 
mainly presenting spectral density function results.  
 In particular, we analize the main changes in $\tilde{A}(\vec{k},\omega)$ with doping, fixing $\vec{k}$ at the relevant high-symmetry points of the BZ which were 
  experimentally studied. We compare with  ARPES[76]
  and angle integrated valence band photoemission experiments[77] on undoped, Co-doped (electron doped), and K-doped (hole doped)  Ba-122 single crystals,
    in which asymmetric electron-hole doping effects on the electronic structure were reported,[76,77] 
    confirming previous indications of Hall and transport experiments.[79] 
   
      \begin{figure}[h!]
  \begin{center}
  \includegraphics[width=8.6cm]{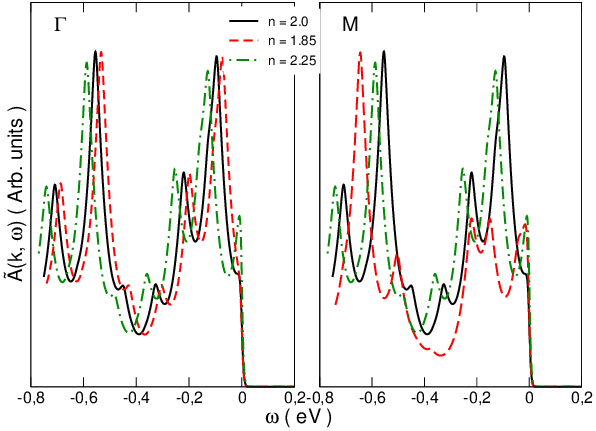}
    \caption[]{Comparison of the effect of electron vs. hole doping on $ \tilde{A} (k, \omega)$, at $\Gamma$ and $M$ points. T = 40 K,  
    $U = V = 3.50 eV$, $\nu= 9$. Other parameters as in Fig.1.}
  \label{doping2}
  \end{center}
  \end{figure}
  
  In Fig. \ref{doping2}, we show $\tilde{A}(\vec{k},\omega)$ for three cases, corresponding to the fillings of the Ba-122 compounds 
  studied in Refs.[76,77]: the undoped ($n=2$), hole-doped ($n=1.85$), and electron-doped ($n=2.25$) systems. 
   at the $\Gamma$ and $M$ Brillouin zone points. As observed at $\Gamma$ with ARPES,[77] 
    we obtain two main peaks (apart from a few other peaks all of lower intensity):  
   one of them centered at binding energy 0.55 eV  for the undoped compound, and the other one near the Fermi level,  
   which shift their positions according to doping  following the trend observed in experiments. 
   Concretely, the peak  at $\omega \sim -0.55$ eV for the undoped system, appears shifted away from the Fermi level: at $\omega \sim -0.60$ eV 
   for the electron-doped system,  while for the hole-doped system it appears shifted in the opposite direction: 
    at $\omega \sim -0.50$ eV. In fact, at $\Gamma$ the whole spectrum appears shifted analogously as a function of doping. 
   Thus, electron-hole  asymmetric effects  such  as experimentally observed[76,77]  are present in the renormalized electronic structure of our model.
   Regarding the spectral densities at $M$ shown in  Fig. \ref{doping2}, the situation is seen to differ: 
   though the same trend of spectral weight shifts upon doping sign is seen for the part of the spectrum corresponding to energies closer to the Fermi level 
   ($ \omega > - 0.4$ eV), an important spectral weight redistribution is also present upon doping. 

 \begin{figure}[t!]
  \begin{center}
\includegraphics[width=8.6cm]{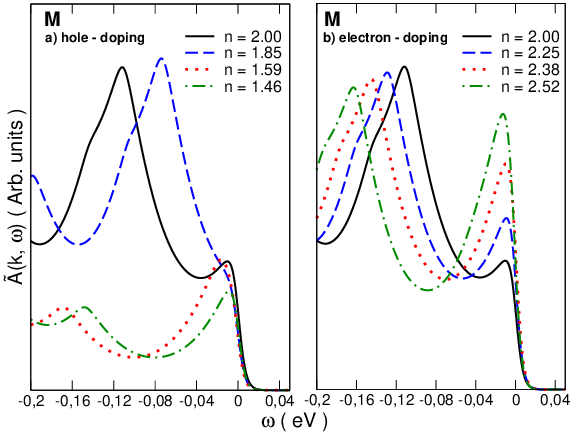}
    \caption[]{ Effect of a) hole doping and b) electron doping on $ \tilde{A} (k, \omega)$, at $M$ point. T = 40 K, U = V = 3.50 eV. 
    $\nu= 9$. Other parameters as in Fig.1.}
  \label{doping3}
  \end{center}
  \end{figure}

  \begin{figure}[t]
  \begin{center}
  \includegraphics[width=8.6cm]{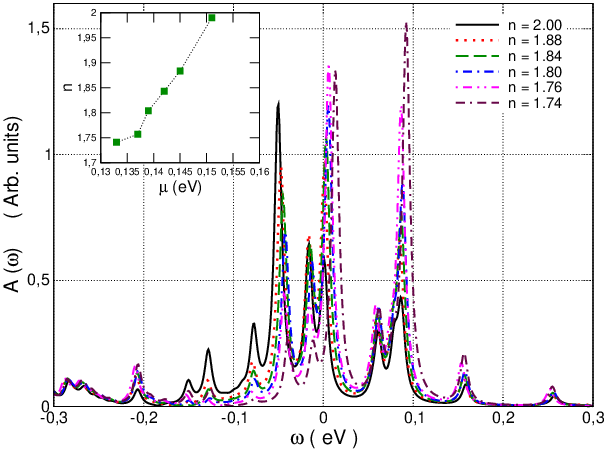}
    \caption[]{ Total DOS $A (\omega)$, for different values of  \textbf{hole} doping (indicated in the figure) at T = 20 K,  
    $U = V = 3.50 eV$.     Other parameters as in Fig.1. Inset:  dependence of the chemical potential on total band filling. }  
  \label{hole-doping}
  \end{center}
  \end{figure}

\nopagebreak
 \begin{figure*}[t]
\vspace{0.5cm}
 \begin{center}
  \includegraphics[]{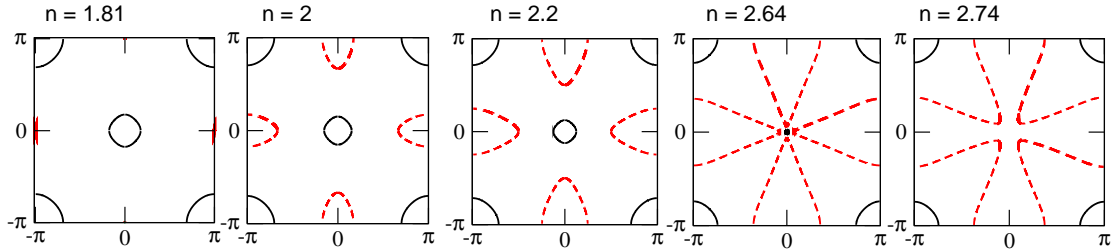}
 \caption{Correlated two-orbital model: doping evolution of the  (large) Fermi surface (one Fe atom / cell), at $T = 40$ K.
   Electron- (dashed) and hole pockets (solid lines), at five  indicated fillings  
  ($ n = 2 $: parent compound).  Parameters: $ U = V= 3.5 eV$, tight-binding parameters from Ref.[30]}
  \end{center}
\end{figure*}    
  
  In Fig. \ref{doping3} we focus on the spectral density at M in the range of energies closer to the  Fermi level  studied by ARPES in Ref.[76], 
and exhibit our results for more values of charge doping: (a) for hole doping, and in (b) for electron-doping.  
The nature of the spectral weight redistribution that we find takes place upon doping is in qualitative agreement with the reported data.[76]
Increasing hole-doping, spectral weight from lower energies is transferred towards the Fermi level. 
Meanwhile, increasing electron doping we obtain a smooth increase  of the quasiparticle weight at the Fermi level,  
while other states are pushed away to lower energies. 

  In Figure \ref{hole-doping} we complement the results presented in Fig. 3,  showing the total DOS
   for a larger number of hole doping values 
  One clearly appreciates that  increasing hole-doping,  a rigid-band-like shift of all peaks towards
the Fermi level is obtained, thus increasing spectral weight there in agreement with ARPES results on doped BaFe$_{2}$As$_{2}$,[76]
where a smooth chemical potential shift upon doping, related to the effective masses of the low-energy valence states, was observed. 
The filling-dependence of the chemical potential in our model is shown as an inset in Fig. \ref{hole-doping}.

  Finally, in Figure C.8 we exhibit  effects of doping  on the renormalized Fermi surface (FS) topology  of the correlated two-orbital model. Notice that 
  several Lifshitz transitions are obtained. Starting from the FS typical of parent compounds ($ n = 2 $), doping with electrons mainly reduces the hole point around 
  the BZ centre $\Gamma$, until it disappears at critical electron filling $ n_{e} \sim 2.65$. At the same time, one can see that electron doping increases 
  the four electron pockets located around the BZ points: X,Y, -X,-Y, until a new FS topology emerges at  $ n_{e}$, with a larger electron FS around $\Gamma$. 
  This is qualitatively consistent with experimental reports [22,23].   Similarly, notice in Figure C.8 that hole doping reduces the four above-mentioned electron pockets 
  until at critical hole filling $ n_{h} \sim 1.81$ they disappear, and a new FS topology emerges: only consisting of the hole pockets.  


\subsection{Temperature dependence of the electronic structure: effect of doping.}

After having discussed in section 3.4  the effects of temperature on the electronic structure for the parent compounds ( n = 2 ), 
here we compare them  with the results predicted for doped systems. 

In Figure C.9  we focus on the  total density of states:   the results shown evidentiate that   the non-trivial temperature dependent renormalization effects by correlations 
 found for the parent  compounds,  shown in Fig. C.9.(b) and previously discussed in connection  with Figure 4, 
 are greatly reduced when doping is introduced.  In particular, Fig.C.9.(a)  shows that for hole doping,  
 increasing temperature,  the dominant peak near the Fermi level   is not displaced   and the main effect of  temperature 
 is the usual expected thermal broadening (and  consequent decrease of the height of the peak). While for electron-doping, 
 Fig.C.9.(c) shows that a certain non-trivial redistribution of spectral weight still appears with temperature, though much less 
 important than for the parent compounds: notice that the dominant peak near the Fermi level, 
  is shifted to lower energies by an increase of temperature, reducing its height.

   \begin{figure}[h!]
  \begin{center}
   \includegraphics[width=8.6cm]{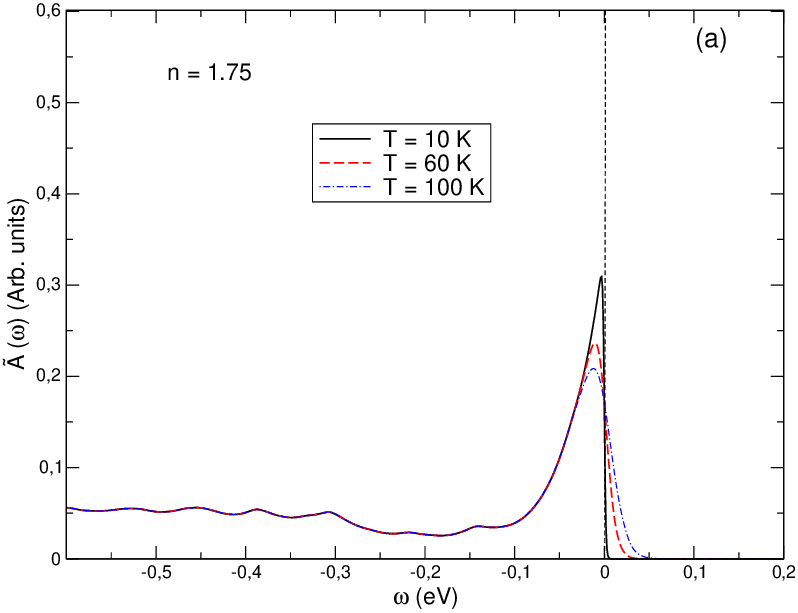}
  \includegraphics[width=8.6cm]{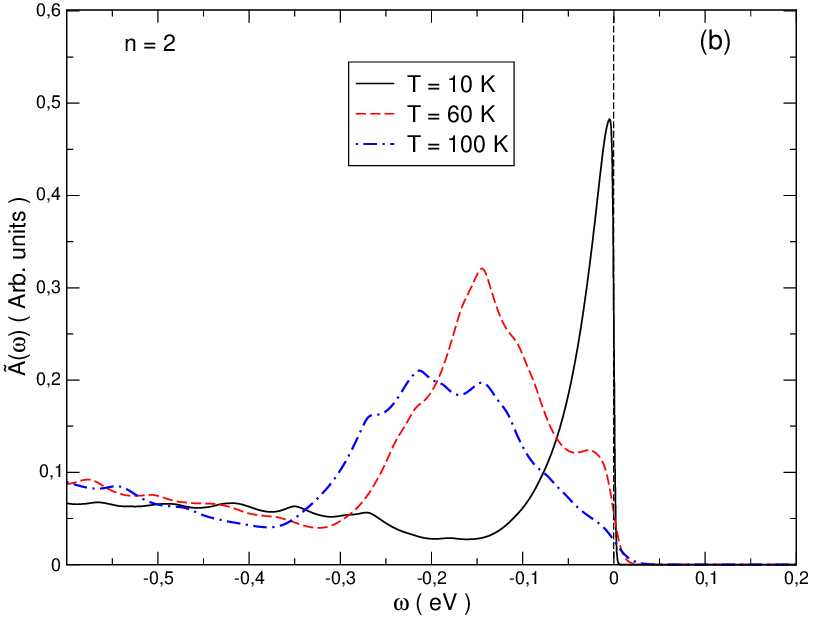}
  \includegraphics[width=8.6cm]{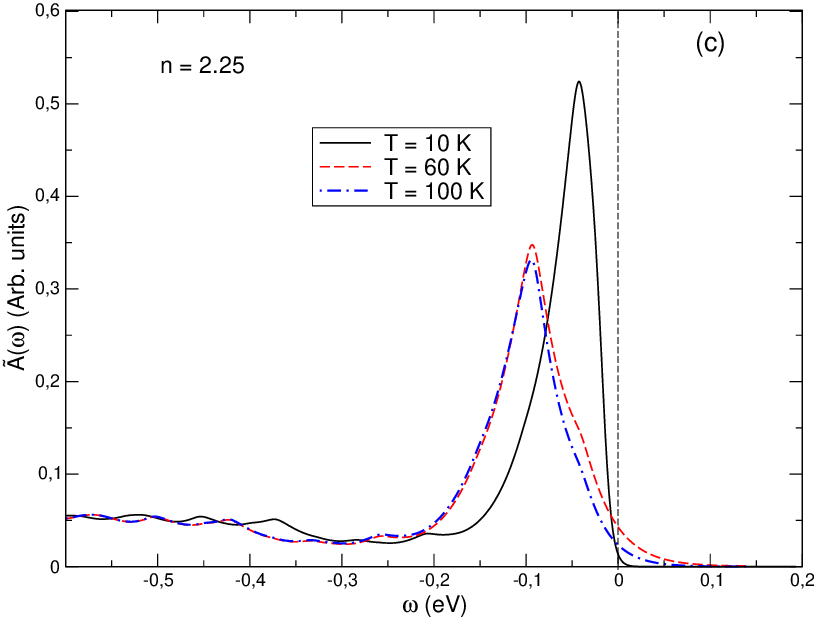}
     \caption[]{Total DOS  $ \tilde{A} (\omega)$  shown at three temperatures: T = 10 K, 60 K and 100 K, 
     for fillings:
      (a) $ n = 1.75 $  corresponding to hole-doping; 
     (b) $ n = 2 $ ( parent compounds); 
     (c) $ n = 2.25 $ corresponding to electron-doping. 
    $ U = V = 3.5 eV$, $\nu= 8$.    Other parameters as in Fig.1.}
  \label{suppl-fig9}
  \end{center}
  \end{figure}  

We also explored the effect of doping on the temperature dependent spectral density function results. 
In Figure C.10, for an electron-doped system, 
 we focus on the  spectral density function:  in particular for the three Brillouin zone points previously 
discussed in connection with Figure 6, exemplifying BZ points we could identify for the parent compounds 
where larger non-trivial temperature dependent renormalization effects by correlations are predicted.  
Notice that doping strongly affects those k-dependent results:
 for the two BZ points  shown in Figs. C.10.(a) and C.10.(c),  the presence of doping 
 has suppressed the temperature dependence. While Fig. C.10.(b) shows that 
 for $\vec{k_2}=(0.6\pi,0.6\pi)$ some temperature dependence is  still retained. For hole doping, 
 we found that no relevant temperature dependent effects are retained at these three BZ points. 

   \begin{figure}[h!]
  \begin{center}
   \includegraphics[width=8.6cm]{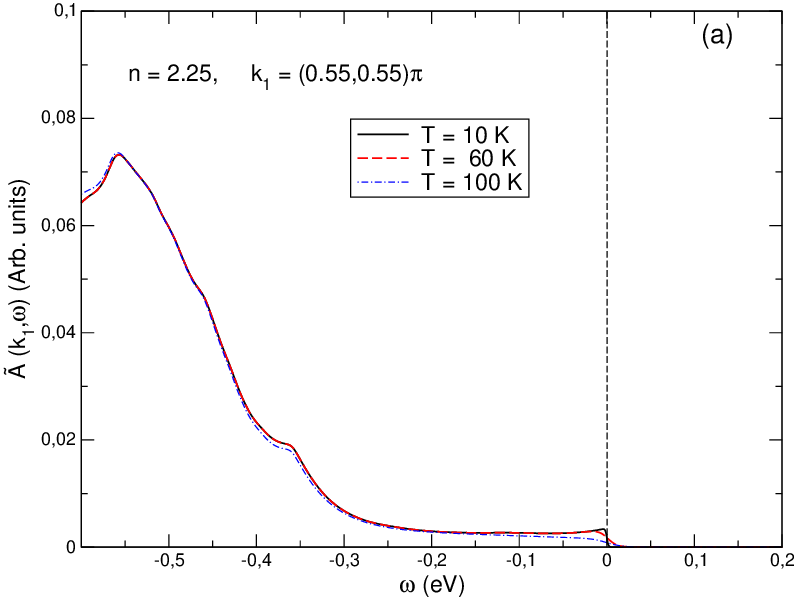}
  \includegraphics[width=8.6cm]{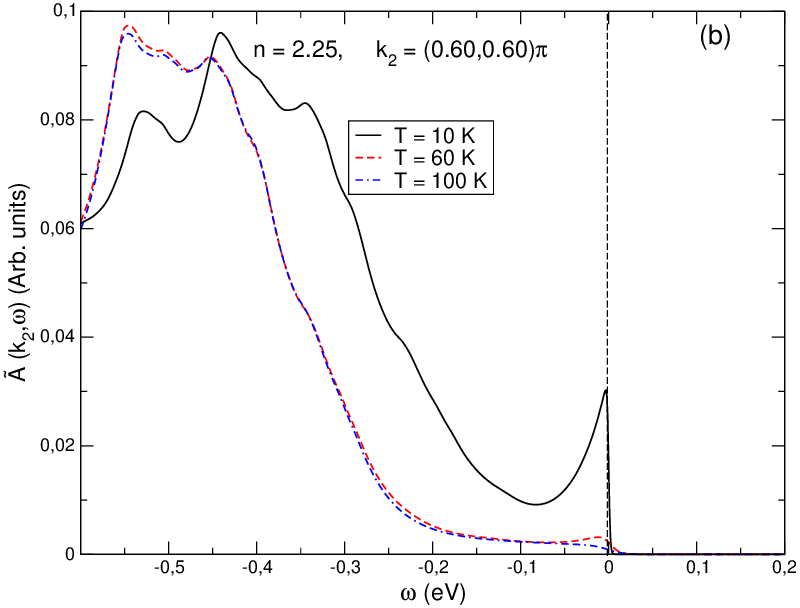}
  \includegraphics[width=8.6cm]{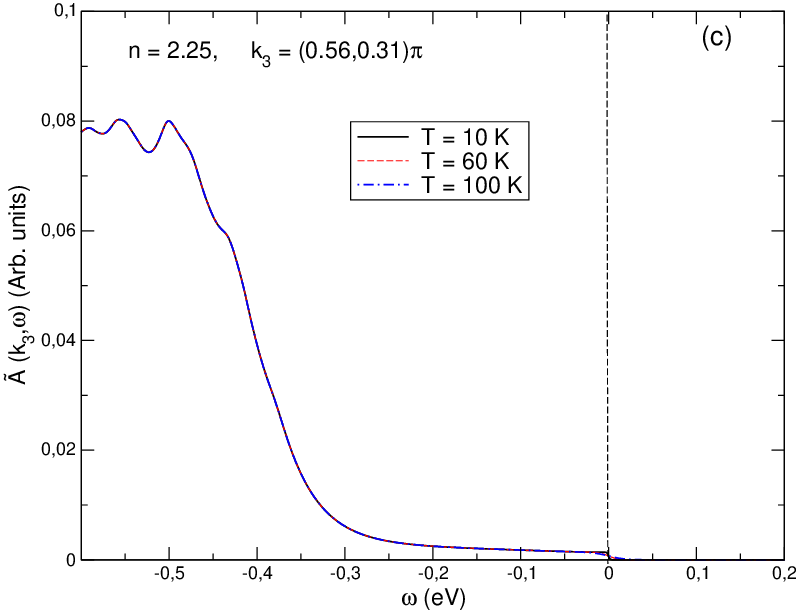}
     \caption[]{ Temperature dependence of $ \tilde{A} (k, \omega)$ for  $ n= 2.25 $: shown  at three BZ points,  respectively indicated in each plot. 
     Parameters:  U = V = 3.5 eV , $\nu= 8$.  Other parameters as in Fig.1.}
  \label{suppl-fig10}
  \end{center}
  \end{figure}

\end{document}